\documentclass[fleqn,usenatbib]{mnras}
\usepackage[utf8]{inputenc}
\usepackage[T1]{fontenc} 
\PassOptionsToPackage{fleqn}{amsmath}
\PassOptionsToPackage{british}{babel}
\usepackage{fancyhdr}
\usepackage{amsmath}
\usepackage{amssymb}
\usepackage{graphicx}
\usepackage{xargs}[2008/03/08]
\usepackage{makecell}
\usepackage{bm}

\makeatletter
\newcommand*{\centerfloat}{%
  \parindent \z@
  \leftskip \z@ \@plus 1fil \@minus \textwidth
  \rightskip\leftskip
  \parfillskip \z@skip}
\makeatother

\makeatletter

\title[Circumbinary-disc populations]{Circumbinary discs for stellar population models}

\author[Izzard \& Jermyn]{ Robert G. Izzard$^{1,2}$ and Adam S. Jermyn$^{2,3}$\\
$^{1}$Astrophysics group, Department of Physics, University of Surrey,
Guildford, GU2 7XH, United Kingdom.\\
$^{2}$Institute of Astronomy, Madingley Road, Cambridge, CB3 0HA,
United Kingdom.\\
$^{3}$Center for Computational Astrophysics, Flatiron Institute, New York, NY 10010, USA.}
\providecommand{\tabularnewline}{\\}
\usepackage[british]{babel}         
\usepackage{newtxtext}               
\usepackage[slantedGreek]{newtxmath}  
 
\usepackage[T1]{fontenc}   
\usepackage{graphicx}        
\usepackage[fleqn]{amsmath}
\DeclareMathSymbol{,}{\mathpunct}{operators}{"2C}
\usepackage{mleftright}
\renewcommand{\left}{\mleft}
\renewcommand{\right}{\mright}

\let\jnl@style=\rmfamily 
\def\ref@jnl#1{{\jnl@style#1}}%
\providecommand\aj{\ref@jnl{AJ}}
\providecommand\araa{\ref@jnl{ARA\&A}}
\providecommand\apj{\ref@jnl{ApJ}}
\providecommand\apjl{\ref@jnl{ApJL}}     
\providecommand\apjs{\ref@jnl{ApJS}}
\providecommand\ao{\ref@jnl{ApOpt}}
\providecommand\apss{\ref@jnl{Ap\&SS}}
\providecommand\aap{\ref@jnl{A\&A}}
\providecommand\aapr{\ref@jnl{A\&A~Rv}}
\providecommand\aaps{\ref@jnl{A\&AS}}
\providecommand\azh{\ref@jnl{AZh}}
\providecommand\baas{\ref@jnl{BAAS}}
\providecommand\icarus{\ref@jnl{Icarus}}
\providecommand\jrasc{\ref@jnl{JRASC}}
\providecommand\memras{\ref@jnl{MmRAS}}
\providecommand\mnras{\ref@jnl{MNRAS}}
\providecommand\pra{\ref@jnl{PhRvA}}
\providecommand\prb{\ref@jnl{PhRvB}}
\providecommand\prc{\ref@jnl{PhRvC}}
\providecommand\prd{\ref@jnl{PhRvD}}
\providecommand\pre{\ref@jnl{PhRvE}}
\providecommand\prl{\ref@jnl{PhRvL}}
\providecommand\pasp{\ref@jnl{PASP}}
\providecommand\pasj{\ref@jnl{PASJ}}
\providecommand\qjras{\ref@jnl{QJRAS}}
\providecommand\skytel{\ref@jnl{S\&T}}
\providecommand\solphys{\ref@jnl{SoPh}}
\providecommand\sovast{\ref@jnl{Soviet~Ast.}}
\providecommand\ssr{\ref@jnl{SSRv}}
\providecommand\zap{\ref@jnl{ZA}}
\providecommand\nat{\ref@jnl{Nature}}
\providecommand\iaucirc{\ref@jnl{IAUC}}
\providecommand\aplett{\ref@jnl{Astrophys.~Lett.}}
\providecommand\apspr{\ref@jnl{Astrophys.~Space~Phys.~Res.}}
\providecommand\bain{\ref@jnl{BAN}}
\providecommand\fcp{\ref@jnl{FCPh}}
\providecommand\gca{\ref@jnl{GeoCoA}}
\providecommand\grl{\ref@jnl{Geophys.~Res.~Lett.}}
\providecommand\jcp{\ref@jnl{JChPh}}
\providecommand\jgr{\ref@jnl{J.~Geophys.~Res.}}
\providecommand\jqsrt{\ref@jnl{JQSRT}}
\providecommand\memsai{\ref@jnl{MmSAI}}
\providecommand\nphysa{\ref@jnl{NuPhA}}
\providecommand\physrep{\ref@jnl{PhR}}
\providecommand\physscr{\ref@jnl{PhyS}}
\providecommand\planss{\ref@jnl{Planet.~Space~Sci.}}
\providecommand\procspie{\ref@jnl{Proc.~SPIE}}

\providecommand\actaa{\ref@jnl{AcA}}
\providecommand\caa{\ref@jnl{ChA\&A}}
\providecommand\cjaa{\ref@jnl{ChJA\&A}}
\providecommand\jcap{\ref@jnl{JCAP}}
\providecommand\na{\ref@jnl{NewA}}
\providecommand\nar{\ref@jnl{NewAR}}
\providecommand\pasa{\ref@jnl{PASA}}
\providecommand\rmxaa{\ref@jnl{RMxAA}}

\providecommand\maps{\ref@jnl{M\&PS}}
\providecommand\aas{\ref@jnl{AAS Meeting Abstracts}}
\providecommand\dps{\ref@jnl{AAS/DPS Meeting Abstracts}}

\providecommand\cac{\ref@jnl{Computational Astrophysics and Cosmology}}

\usepackage[export]{adjustbox}
\usepackage{xcolor}

\usepackage[british]{babel}             
\usepackage{newtxtext}                  
\usepackage[slantedGreek]{newtxmath}    

\usepackage[T1]{fontenc}
\usepackage{aecompl}
\usepackage{mleftright}
\usepackage{makecell}
\usepackage{rotating}

 \hypersetup{pdfauthor={R.G. Izzard and A.S. Jermyn},
               pdftitle={Circumbinary Discs},
               pdfkeywords={binary stars, discs},
               bookmarksnumbered=true}

\makeatother
\newcommand\eqq{Eq.}
\newcommand\secref{Sec.}
\newcommand\figref{Fig.}

\usepackage{babel}

\begin{document}
\maketitle
\global\long\def\sun{\odot}%

\global\long\def\Teff{T_{\mathrm{eff}}}%

\global\long\def\logg{\log_{10}\left(g/\mathrm{cm}^{2}\mathrm{s}^{-1}\right)}%

\global\long\def\loggsquare{\log_{10}\left[g/\mathrm{cm}^{2}\mathrm{s}^{-1}\right]}%

\global\long\def\Zsol{Z_{\sun}}%

\newcommandx\squarebracket[2][usedefault, addprefix=\global, 1=, 2=]{\left[\mathrm{#1}/\mathrm{#2}\right]}%

\global\long\def\squarebracketdefinition#1#2{\log_{10}\left(N_{\mathrm{#1}}/N_{\mathrm{#1},\odot}\right)-\log_{10}\left(N_{\mathrm{#2}}/N_{\mathrm{#2},\odot}\right)}%

\global\long\def\aFe{\squarebracket[\alpha][Fe]}%

\global\long\def\CFe{\squarebracket[C][Fe]}%

\global\long\def\CN{\squarebracket[C][N]}%

\global\long\def\FeH{\squarebracket[Fe][H]}%

\global\long\def\CNdef{\squarebracketdefinition CN}%

\global\long\def\FeHdef{\squarebracketdefinition{Fe}H}%

\global\long\def\binaryc{\code{binary\_c}}%

\global\long\def\SSE{\code{sse}}%

\global\long\def\BSE{\code{bse}}%

\global\long\def\STARS{\code{stars}}%

\global\long\def\TWIN{\code{twin}}%

\global\long\def\ROSE{\code{rose}}%

\global\long\def\MESA{\code{mesa}}%

\global\long\def\BEC{\code{bec}}%
\global\long\def\emcee{\code{emcee}}%

\global\long\def\mdisc{M_{\mathrm{disc}}}%
\global\long\def\jdisc{J_{\mathrm{disc}}}%

\global\long\def\mbin{\left(M_{1}+M_{2}\right)}%

\global\long\def\kb{k_{\mathrm{B}}}%

\global\long\def\oderiv#1{\mathrm{d}#1}%
\global\long\def\pderiv#1{\partial#1}%

\global\long\def\pd#1#2{\frac{\pderiv{#1}}{\pderiv{#2}}}%
\global\long\def\pdl#1#2{\pderiv{#1}/\pderiv{#2}}%

\global\long\def\d#1#2{\frac{\oderiv{#1}}{\oderiv{#2}}}%
\global\long\def\dl#1#2{\oderiv{#1}/\oderiv{#2}}%

\global\long\def\dc#1#2#3{\left(\d{#1}{#2}\right)_{#3}}%
\global\long\def\dlc#1#2#3{\left(\dl{#1}{#2}\right)#3}%

\global\long\def\pdt#1#2{\pd{^{2}#1}{#2^{2}}}%
\global\long\def\pdtl#1#2{\pdl{^{2}#1}{#2^{2}}}%

\global\long\def\dt#1#2{\d{^{2}#1}{#2^{2}}}%
\global\long\def\dtl#1#2{\dl{^{2}#1}{#2^{2}}}%

\global\long\def\dtc#1#2#3{\left(\dt{#1}{#2}\right)_{#3}}%
\global\long\def\dtlc#1#2#3{\left(\dtl{#1}{#2}\right)_{#3}}%

\global\long\def\pdc#1#2#3{\left(\pd{#1}{#2}\right)_{#3}}%
\global\long\def\pdcl#1#2#3{\left(\pdl{#1}{#2}\right)_{#3}}%

\global\long\def\pdtc#1#2#3{\left(\pdt{#1}{#2}\right)_{#3}}%
\global\long\def\pdtcl#1#2#3{\left(\pdtl{#1}{#2}\right)_{#3}}%

\global\long\def\integral#1#2#3#4{\int_{#3}^{#4}#1\,\oderiv{#2}}%
\global\long\def\lineintegral#1#2#3#4{\varoint_{#3}^{#4}#1\,\oderiv{#2}}%

\global\long\def\answer#1{{\color{red}\underline{\underline{#1}}}}%

\global\long\def\vvec#1#2#3{\left(\begin{array}{c}
#1\\
#2\\
#3
\end{array}\right)}%

\global\long\def\hvec#1#2#3{\left(#1,\,#2,\,#3\right)}%

\global\long\def\vector#1{\bm{#1}}%

\global\long\def\matrix#1{\bm{#1}}%
\global\long\def\m#1{\bm{#1}}%

\global\long\def\dvec#1#2{\left(\begin{array}{c}
#1\\
#2
\end{array}\right)}%
\global\long\def\dmatrix#1#2#3#4{\left(\begin{array}{cc}
#1 & #2\\
#3 & #4
\end{array}\right)}%

\global\long\def\tvec#1#2#3{\left(\begin{array}{c}
#1\\
#2\\
#3
\end{array}\right)}%
\global\long\def\tmatrix#1#2#3#4#5#6#7#8#9{\left(\begin{array}{ccc}
#1 & #2 & #3\\
#4 & #5 & #6\\
#7 & #8 & #9
\end{array}\right)}%

\global\long\def\dmatrix#1#2#3#4{\left(\begin{array}{cc}
#1 & #2\\
#3 & #4
\end{array}\right)}%
\global\long\def\dhvec#1#2{\left(\begin{array}{cc}
#1 & #2\end{array}\right)}%

\global\long\def\hyphen{\text{--}}%

\global\long\def\deriv{\mathrm{d}}%

\global\long\def\e{\mathrm{e}}%

\global\long\def\i{\mathrm{i}}%

\global\long\def\crossproduct{\bm{\times}}%

\global\long\def\dotproduct{\bm{\cdot}}%

\global\long\def\bfnabla{~\nabla}%

\global\long\def\dotproduct{\bm{\cdot}}%

\global\long\def\Tr#1{\mathrm{Tr}\left(#1\right)}%

\global\long\def\Real{\mathbb{R}}%

\global\long\def\Poin{\text{Poincar\ensuremath{\mathrm{\'e}}}}%

\global\long\def\PBTh{\text{Poincar\ensuremath{\mathrm{\'e}}-Bendixson}}%

\global\long\def\sgn{\mathrm{sgn}}%
\global\long\def\order#1{{\cal O}\left(#1\right)}%

\global\long\def\f#1#2{#1\left(#2\right)}%

\global\long\def\Ra{\mathrm{Ra}}%

\global\long\def\pc{\mathrm{\,per\,cent}}%

\global\long\def\msun{\mathrm{M}_{\odot}}%

\global\long\def\rsun{\mathrm{R}_{\odot}}%

\global\long\def\lsun{\mathrm{L}_{\odot}}%

\global\long\def\sun{\odot}%

\global\long\def\code#1{\normalfont\textsc{{#1}}}%

\global\long\def\millisecond{\mathrm{ms}}%

\global\long\def\second{\mathrm{s}}%

\global\long\def\minute{\mathrm{min}}%

\global\long\def\hour{\mathrm{h}}%

\global\long\def\days{\mathrm{d}}%

\global\long\def\yr{\mathrm{yr}}%

\global\long\def\Myr{\mathrm{Myr}}%

\global\long\def\Gyr{\mathrm{Gyr}}%

\global\long\def\magnitude{\mathrm{mag}}%

\global\long\def\AU{\mathrm{au}}%

\global\long\def\cm{\mathrm{cm}}%

\global\long\def\m{\mathrm{m}}%

\global\long\def\km{\mathrm{km}}%

\global\long\def\junits{\mathrm{g}\,\mathrm{cm}^{2}\,\mathrm{s}^{-1}}%

\global\long\def\IRAS{\text{IRAS\,08544-4431}}%
\global\long\def\flux{F_{\star}}%
\global\long\def\rr{\frac{R}{R_{\star}}}%
\global\long\def\rin{R_{\mathrm{in}}}%
\global\long\def\rout{R_{\mathrm{out}}}%
\global\long\def\Mdisc{M_{\mathrm{disc}}}%
\global\long\def\Jdisc{J_{\mathrm{disc}}}%
\global\long\def\Ldisc{L_{\mathrm{disc}}}%
\global\long\def\Jorb{J_{\mathrm{orb}}}%
\global\long\def\Mdot#1{\dot{M}_{\mathrm{#1}}}%
\global\long\def\Mdotdisc{\Mdot{disc}}%
\global\long\def\Mdotvisc{\Mdot{visc}}%
\global\long\def\Jdot#1{\dot{J}_{\mathrm{#1}}}%
\global\long\def\Jdotdisc{\Jdot{disc}}%
\global\long\def\sigmaz{\Sigma_{0}}%
\global\long\def\Mmin{M_{\mathrm{min}}}%
\global\long\def\Mmax{M_{\mathrm{max}}}%
\global\long\def\Lmin{L_{\mathrm{min}}}%
\global\long\def\Lmax{L_{\mathrm{max}}}%
\global\long\def\fring{f_{\mathrm{Ring}}}%
\global\long\def\ace{\alpha_{\mathrm{CE}}}%
\global\long\def\lce{\lambda_{\mathrm{CE}}}%
\global\long\def\ftid{f_{\mathrm{tid}}}%
\global\long\def\hin{H_{\mathrm{in}}}%
\global\long\def\hout{H_{\mathrm{out}}}%
\global\long\def\MCE{M_{\mathrm{CE}}}%
\global\long\def\JCE{J_{\mathrm{CE}}}%
\global\long\def\tin{T_{\mathrm{in}}}%
\global\long\def\tout{T_{\mathrm{out}}}%
\global\long\def\adisc{\alpha_{\mathrm{disc}}}%
\global\long\def\fvisc{f_{\mathrm{visc}}}%
\global\long\def\fviscJ{f_{\mathrm{visc,}J}}%
\global\long\def\fres{f_{\mathrm{res}}}%
\global\long\def\fX{f_{\mathrm{X}}}%
\global\long\def\fM{f_{\mathrm{M}}}%
\global\long\def\fJ{f_{\mathrm{J}}}%
\global\long\def\epostCE{e_{\mathrm{post-CE}}}%
\global\long\def\emax{e_{\mathrm{max}}}%
\global\long\def\tdisc{t_{\mathrm{disc}}}%
\global\long\def\cs{c_{\mathrm{s}}}%
\global\long\def\abin{a_{\mathrm{bin}}}%
\global\long\def\Pbin{P_{\mathrm{bin}}}%
\global\long\def\jcgs{\junits}%
\global\long\def\kappaunits{\mathrm{cm}^{2}\,\mathrm{g^{-1}}}%

\begin{abstract}

We develop a rapid algorithm for the evolution of stable, circular, circumbinary discs suitable for parameter estimation and population synthesis modelling. Our model includes disc mass and angular momentum changes, accretion onto the binary stars, and binary orbital eccentricity pumping. We fit our model to the {post-asymptotic giant branch (post-AGB)} circumbinary disc around \IRAS{} finding reasonable agreement despite the simplicity of our model.  Our best-fit disc has a mass of about $0.01\,\msun$ and angular momentum $2.7\times10^{52}\,\junits \simeq 9 \msun\,\mathrm{km}\,\mathrm{s}^{-1}\,\mathrm{au}$, corresponding to $0.0079$ and $0.16$ of the common-envelope mass and angular momentum respectively. The best-fit disc viscosity is $\adisc=5\times10^{-3}$ and our tidal torque algorithm can be constrained such that the inner edge of the disc $\rin \sim 2a$. The inner binary eccentricity reaches about $0.13$ in our best-fitting model of \IRAS{}, short of the observed $0.22$.

The circumbinary disc evaporates quickly when the post-AGB star reaches a temperature of $\sim\! 6\times10^4\,\mathrm{K}$, suggesting that planetismals must form in the disc in about $10^{4}\,\yr$ if secondary planet formation is to occur, while accretion from the disc onto the stars at $\sim\! 10$ times the inner-edge viscous rate can double the disc lifetime.
\end{abstract}



\section{Introduction}

\label{sec:Introduction}
Understanding the evolution of binary stars
is vital to many areas of modern astrophysics. Phenomena that occur
uniquely or primarily in binary systems include type Ia supernovae,
X-ray binaries, thermonuclear novae, X-ray bursts, stellar rejuvenation
and blue stragglers, the formation of stripped stars such as Wolf-Rayet
and sdB/O stars, and the merging of stars with associated signals
in gravitational waves and as kilonovae. Binary stars are thus fundamental
to astrophysics, yet many aspects of their evolution remain poorly
understood \citep{2017PASA...34....1D}.

Mass transfer is a vital process in binary-star evolution, and while
Roche-lobe overflow as a basic concept is reasonably understood, quantitative
details regarding efficiency of mass transfer are not yet pinned down.
Transferred material can be retained by the companion star or lost
from the system, and the fate of associated angular momentum is equally
uncertain. Because the specific angular momentum of transferred material
is often large compared to the associated stellar specific angular
momentum, discs form in many mass-transferring binaries. These discs
can be either circumstellar, around one or both stars, or circumbinary,
around the binary as a whole. From observations we know that, at least
in some binaries, a disc forms outside the binary orbit in a circumbinary
disc \citep{2016A&A...596A..92K,2018MNRAS.473..317H}. The circumbinary disc material may have been ejected by a wind
from its parent star, or during non-conservative mass transfer or
common-envelope ejection \citep[e.g.][]{2011MNRAS.417.1466K}. A better understanding of the formation
and evolution of such discs, and their interaction with their binary
parent stars, is the aim of this paper.

Binary {post-giant branch (post-GB) and post-asymptotic giant branch (post-AGB) stars (together referred to below as post-(A)GB)} are a class of
systems which are observed to have surrounding discs \citep{2013A&A...557A.104B},
indeed \emph{all} post-AGB systems with discs are binaries. These
binaries contain a relatively hot ($\gtrsim 6\times10^{3}\,\mathrm{K}$), bright, compact object which has
just left the AGB and a usually main-sequence dwarf companion star.
The post-AGB star has a compact degenerate core surrounded by a thin envelope containing
about $10^{-3}\,\msun$ of mostly hydrogen. At the interface of the
core and envelope is a hydrogen-burning shell which generates the
typically $10^{4}\,\lsun$ luminosity of the system.

The circumbinary discs in post-AGB systems are mostly Keplerian and
stable \citep{2006A&A...448..641D,2021A&A...648A..93G}. Their masses are up to $10^{-2}\,\msun$, with outer diameters
of about $10^{5}\,\rsun$ and angular momenta up to $10^{52}\,\junits$
which is similar to their inner binary system. Mass loss is observed
from the outer part of the disc, which can be sub-Keplerian, and inflow
at the inner edge onto the binary stars is observed at rates
up to $10^{-7}\,\msun\,\yr^{-1}$ \citep{2018A&A...614A..58B}. Material
that flows inward is poor in metals because these are lost in a dusty
outflow, thus the surface of the post-AGB star is polluted with low-metallicity
material \citep{2020A&A...642A.234O}. This accretion may also prolong the life of the post-AGB star and circumbinary disc,
 and hence increase the number of post-AGB binaries relative
to the equivalent single stars.

The orbits of post-AGB stars with circumbinary discs have periods
ranging from tens to thousands of days \citep{2018A&A...620A..85O}. While the wide systems, with
periods longer than about $1000\,\mathrm{d}$, may form circumbinary discs from wind
material ejected during the previous AGB phase \citep{2019A&A...629A.103S}, systems with shorter
periods probably passed through a phase of Roche-lobe overflow followed by common-envelope evolution
because their orbits are too small to contain an AGB star of radius $\sim 500\mathrm{\,R_{\odot}}$ \citep{2012IAUS..283...95I,2013A&ARv..21...59I}.
During common-envelope evolution, the AGB core and its companion transfer orbital energy and angular momentum into the envelope and thus eject it.  Evolved AGB stars are poorly bound \citep{2016RAA....16..126W}, so only a little energy is required from the orbit for envelope ejection, and hence the orbit also shrinks little during this phase. 

The circumbinary disc interacts with its binary system through both
mass accretion and resonant interactions that may increase the eccentricity of the binary orbit.
Post-AGB systems with periods shorter than $1000\,\days$
should be circular because of efficient tidal dissipation, and yet
many are observed to have eccentricities up to $e=0.4$, and a few
have $e\gtrsim 0.6$ \citep{2018A&A...620A..85O}. Resonant interactions have been explored as sources
of this eccentricity previously, but the models employed were either
rather limited in that they assumed a fixed disc mass and lifetime
\citep{2013A&A...551A..50D}, or did not include feedback with stellar
evolution and subsequent disc evaporation \citep{2016ApJ...830....8R}.
The magnitude of eccentricity pumping depends crucially on both the
disc lifetime and mass.

Recent interest in circumbinary planets, of which about a dozen are now known \citep[e.g.][]{2015A&A...581A..20K,2017MNRAS.468.2932G}, naturally leads to speculation that these may form in evolved stellar systems with circumbinary discs.
Post-AGB binaries are prime candidates for such planet formation given that they host apparently stable circumbinary discs.
Understanding post-AGB circumbinary discs is thus key to also knowing whether they are important sources of planets or other rocky objects that may form debris discs around their offspring \citep{2019MNRAS.489.3896C,2019MNRAS.488.3588Y}.
Many other close-binary formation channels may also be influenced by circumbinary discs. For example, \citet{2019ApJ...876L..11C} investigate the influence of a circumbinary disc on the formation and evolution of X-ray binaries originating from red supergiant systems.

In the following we develop a circumbinary-disc model that is fast enough to be 
suitable for stellar population synthesis modelling yet incorporates
the physics required to study the interaction of a circumbinary disc
with its inner binary star system and the interstellar medium.   
Our model makes reasonable approximations to facilitate fast computation yet allows predictions of system properties for comparison with observations to pin down uncertain physics.
In \secref{}~2 we develop our circumbinary-disc model.
\secref{}~3 explores interaction between the disc and its inner binary system.
\secref{}~4 models the disc recently observed by the Atacama Large Millimeter Array
(ALMA), $\IRAS$ \citep{2018A&A...614A..58B}.
In \secref{}~5 we discuss relevant uncertainties, and the limitations and successes of our model.
We then conclude. 

\section{Circumbinary disc model}

\label{sec:Circumbinary-disc-model}
We require a fast circumbinary
disc model for application in population synthesis calculations
of millions of binary-star systems covering a many-parameter space \citep{2018arXiv180806883I} and, in this paper, to model the disc in the system \IRAS{}. Given the many uncertainties involved, and our requirement for speed of model execution,
we adopt a semi-analytic model guided by observed post-AGB circumbinary
discs. {Our model is conservative, in that it assumes the discs to be in equilibrium and changing slowly relative to their viscous timescale.}

These circumbinary discs are almost Keplerian \citep{2006A&A...448..641D} and, with observed mass loss
rates of $10^{-7}\,\msun\,\yr$ and masses of about $10^{-2}\,\msun$,
are stable on time-scales of at least $10^{4}-10^{5}\,\yr$. This is long
compared to the viscous time-scale at the inner edge of the disc, which
is typically $10^{3}\,\yr$ or shorter. We thus assume that
viscous spreading of the disc \citep[cf.][]{1974MNRAS.168..603L}
is fast, such that the angular momentum flux in the disc adjusts instantly to changes in the disc structure and binary torque.
This allows us to treat mass loss as just changing the total mass of the disc, rather than having to track precisely from where the mass comes and how the disc spreads to restore equilibrium. {These assumptions, and our approach of using matched power-law solutions (\secref{}~\ref{subsec:Solution-method}), are in the spirit of~\citep{1973A&A....24..337S}, \citet{1993ASPC...36..371P} and~\citet{Haiman2009}, though with a somewhat different physical setup. Our disc is split into matched power-law zones depending on which term in the energy equation (Eq.~\ref{eq:T4eq}) dominates under our assumptions of only gas pressure and constant opacity, whereas in, e.g., \citet{1973A&A....24..337S} and \citet{Haiman2009} the splits are based on the dominant opacity and pressure source.}

Our assumptions are valid at the
inner edge of our discs, where the density is greatest and interaction
with the inner binary is strongest, but often not at the outer edge
as we discuss in \secref{}~\ref{subsec:Model-limitations}. 
We find in practice that mass loss barely affects the structure of
the disc except during its evaporation (\secref{}~\ref{subsec:Mass-loss}), at which point the disc is completely disrupted in a few years.
Such fast evolutionary phases are rarely observed, so our model is
appropriate for comparison with most observed, circumbinary, post-AGB
discs. 

\subsection{Disc formation}

\label{subsec:Disc-formation}We form circumbinary discs in binary
systems when a common envelope is ejected. The disc mass, $\mdisc$, is
a fraction, $\fM$, of the ejected common-envelope mass, $\MCE$,
such that,
\begin{equation}
\mdisc  =  \fM\,\MCE\,.\label{eq:fM}
\end{equation}
The disc is given an angular momentum, $\jdisc$, that is a fraction
of the angular momentum in the common envelope, $\JCE$,
\begin{equation}
\Jdisc = \fJ\,\JCE\,.\label{eq:fJ}
\end{equation}
We require $0\leq\fM<1$, to conserve mass, and 
$0\leq\fJ<1$ as we do not expect common-envelope evolution to leave a retrograde central binary.
Typically $\fM\ll1$ and $\fJ\ll1$, although we do not additionally require $\fM=\fJ$.
While we are theoretically free to choose $\fJ$ thus, too much or too little angular momentum leads to immediate partial or full evaporation of the disc (\secref{}~\ref{subsec:Mass-loss}). Such objects are not candidate post-AGB discs.
We also require that $\Mdisc\ll M_{1}+M_{2}$, where $M_{1}$ and
$M_{2}$ are the masses of the stars in the binary system, such that
the disc is gravitationally stable.
Given the observed masses of post-AGB discs are about $0.01\,\msun$ or less,
and that post-AGB binaries have masses of about $1\,\msun$, this
is never a problem.

As an example, consider a circular binary system with,
prior to mass transfer, $M_{1}=1.5\,\msun$, $M_{2}=0.9\,\msun$
and orbital separation $\abin=800\,\rsun$ corresponding to an orbital
period $\Pbin\approx1700\,\days$. The orbital separation is chosen
such that Roche-lobe overflow starts during the {thermally-pulsing asymptotic giant branch (TPAGB)} and leads to
common-envelope evolution. The orbital angular momentum at this time
is about $1.3\times10^{53}\,\jcgs$. Assuming efficient common-envelope
ejection, about $0.6\mathrm{\,M_{\odot}}$ is ejected, so with
$\fM=1.8\pc$ and $\fJ=15\pc$ we have a disc with mass $\mdisc=0.013\,\msun$
and angular momentum $\jdisc=1.25\times10^{52}\,\junits$, similar
to both observed post-AGB discs and the formation scenario proposed by \citet{2011MNRAS.417.1466K}.

\subsection{Disc structure}

\label{subsec:Disc-structure}We assume material in the circumbinary
disc is on circular orbits around the centre of mass of the binary-star system
and is in a Keplerian steady state. We model only gas
in the disc, neglecting dust, because most of the mass in the disc
is expected to be in the gas phase. We assume the disc is thin, i.e.~$H\left(R\right)\ll R$
where $H\left(R\right)$ is the scale height at radius $R$ from the
centre of mass. That we are in a steady state implies we can neglect
time derivatives, and hence the temperature structure of the disc
is given by solutions to the energy conservation equation,
\begin{equation}
\sigma T^{4} = \mathcal{A}+ \mathcal{B}\left(1+\mathcal{C}\right)+\mathcal{D}\,,\label{eq:disc-heat-balance}
\end{equation}
where $T=T(R)$ is the mid-plane temperature of the disc at radius
$R$ and $\sigma$ is the Stefan-Boltzmann constant. \citet{2013ApJ...764..169P}
derive these terms in detail \citep[see also][]{10.1046/j.1365-8711.1998.01652.x}, but briefly they include a viscous heating
term,
\begin{equation}
\mathcal{A} = \frac{27\kappa\nu\Sigma^{2}\Omega^{2}}{64}\,,\label{eq:calA}
\end{equation}
where,
\begin{equation}
\Omega=\Omega\left(R\right) = \sqrt{\frac{G\mbin}{R^{3}}}\,,\label{eq:Omega(R)}
\end{equation}
is the Keplerian angular velocity at radius $R$, $\Sigma=\Sigma(R)$
is the mass column density at radius $R$ (\secref{}~\ref{subsec:Mass-scaling}),
$\kappa=\kappa\left(R\right)$ is the opacity, $\nu=\nu\left(R\right)$
is the kinematic viscosity (\eqq{}~\ref{eq:kinematic-viscosity}) and
$G$ is the gravitational constant. Irradiation from the central star
leads to the heating term,
\begin{equation}
\mathcal{B} = \frac{2\flux}{3\pi}\left(\frac{R_{\star}}{R}\right)^{3}\,,\label{eq:calB}
\end{equation}
where,
\begin{equation}
\flux = \frac{\left(L_{1}+L_{2}\right)}{4\pi R_{\star}^{2}}\,,\label{eq:stellar-flux}
\end{equation}
is the incident luminous stellar flux with corresponding flux-weighted
stellar radius,
\begin{equation}
R_{\star}  =  \frac{F_{1}R_{1}^{3}+F_{2}R_{2}^{3}}{F_{1}R_{1}^{2}+F_{2}R_{2}^{2}}\,,\label{eq:flux-weighed-radius}
\end{equation}
with $F_{i}$ and $R_{i}$ the flux and radius of star $i=1,2$. The
disc geometry is encoded in,
\begin{equation}
\mathcal{C}  =  \frac{3\pi H}{4R_{\star}}\left(\pd{\ln H}{\ln R}-1\right)\,\label{eq:calC}
\end{equation}
and mass loss in,
\begin{equation}
\mathcal{D}  =  \frac{G\mbin}{2\pi a}\dot{M}R^{-2},\label{eq:calD}
\end{equation}
where $a$ is the binary separation and $\dot{M}=\dot{M}\left(R\right)$
is the mass loss rate at radius $R$. 

In the following, we set $\mathcal{D}=0$ and thus ignore mass loss (or
gain) as a sink (or source) of energy in our discs. We test this assumption
in Appendix~\ref{sec:solution-accuracy-checks} and find it to be good within $10\pc$. We are then
left to solve,
\begin{equation}
\sigma T^{4}  =\mathcal{A}+\mathcal{B}\left(1+\mathcal{C}\right),\label{eq:disc-heat-balance-1}
\end{equation}
and we proceed by writing $\mathcal{A}$, $\mathcal{B}$ and $\mathcal{C}$ in
terms of the radius $R$, temperature $T=T(R)$ and a multiplier.
The local sound speed is, assuming an ideal-gas equation of state,
\begin{eqnarray}
\cs\left(R\right)  =  \sqrt{\frac{\Gamma k_{\mathrm{B}}T}{\mu}}\,,
\end{eqnarray}
and the local scale height is, 
\begin{align}
H(R)  =\frac{\cs\left(R\right)}{\Omega(R)}=\sqrt{\frac{\Gamma k_{\mathrm{B}}}{\mu G\mbin}}\,T^{1/2}R^{3/2},\label{eq:H(R)}
\end{align}
where $\Gamma$ is the adiabatic index of the gas in the disc (neutral
hydrogen gas has $\Gamma=1.4$), $k_{\mathrm{B}}$ is the Boltzmann constant
and $\mu$ is the mass of the gas particles in the disc. Differentiating
\eqq{}~\ref{eq:H(R)} by $R$, we find, 
\begin{equation}
\pd{\ln H}{\ln R}  =  \frac{1}{2}\pd{\left(\ln T+3\ln R\right)}{\ln R}=\frac{1}{2}\pd{\ln T}{\ln R}+\frac{3}{2}\,.\label{eq:dlnHdlnR}
\end{equation}
The kinematic viscosity is,
\begin{equation}
\nu(R)  =  \frac{\adisc\cs^{2}}{\Omega}=\frac{\adisc\Gamma k_{\mathrm{B}}T}{\mu}\sqrt{\frac{R^{3}}{G\mbin}}\,,\label{eq:nu(R)}
\end{equation}
where we assume a constant disc viscosity parameter, $\adisc$ {\citep{1973A&A....24..337S}}. Combining
Eqs.~\ref{eq:disc-heat-balance}, \ref{eq:Omega(R)}, \ref{eq:stellar-flux},
\ref{eq:flux-weighed-radius}, \ref{eq:dlnHdlnR} and \ref{eq:nu(R)},
and dividing by $\sigma$ we find,
\begin{equation}
T^{4}  =  a\Sigma^{2}TR^{-3/2}+bT^{1/2}R^{-3/2}+cR^{-3}\,,\label{eq:T4eq}
\end{equation}
where
\begin{equation}
a  =  \frac{27\kappa\adisc\Gamma k_{\mathrm{B}}\sqrt{G\mbin}}{64\sigma\mu}\,,\label{eq:a}
\end{equation}
\begin{equation}
b  =  \frac{F_{\star}R_{\star}^{2}}{4\sigma}\sqrt{\frac{\Gamma k_{\mathrm{B}}}{\mu G\mbin}}\left(1+\pd{\ln T}{\ln R}\right)\approx\frac{F_{\star}R_{\star}^{2}}{4\sigma}\sqrt{\frac{\Gamma k_{\mathrm{B}}}{\mu G\mbin}}\label{eq:b}
\end{equation}
and
\begin{equation}
c  =  \frac{2F_{\star}R_{\star}^{3}}{3\pi\sigma}\,.\label{eq:c}
\end{equation}

\subsection{Solution method}
\label{subsec:Solution-method}
{%
We implement our disc-solution method in the $\binaryc$ code so that disc and binary-system evolution are coupled from the moment of disc formation to its evaporation.
In our model the disc is fully specified by giving its mass and angular momentum.
$\binaryc$ can then reconstruct profiles of quantities in the disc as required.
Hence, we feed $\binaryc$ the initial mass and angular momentum, and then it constructs the disc and evolves it forward in time.
At every $\binaryc$ timestep,} to solve \eqq{}~\ref{eq:T4eq} throughout
the disc we calculate the terms multiplied by $a$, $b$ and $c$,
neglect the smallest two, substitute \eqq{}~\ref{eq:SigmaR-from-flux}
and solve the resulting expression. This gives us a power law relating
temperature and radius, $\log T\propto\log R$.
For these purposes we neglect the term $\partial \ln T / \partial \ln R$ in \eqq{}~\ref{eq:b} because in the region where $\mathcal{B}$ dominates  this evaluates to a constant smaller than unity.

We define zones {of the disc to be the regions in which one of}
the $a$, $b$ or $c$ term{s in} \eqq~\ref{eq:T4eq}
{is largest. In each zone we then neglect the two smaller terms leaving a unqiue} 
temperature-radius power-law relation, $T\left(R\right)$, {in the zone}.
Zones must have radii that satisfy $\rin\leq R\leq\rout$ to be considered part of the disc.

Our calculated $T(R)$ is most erroneous at the edges of the zones. In
the worst case scenario the maximum error in $T^{4}$ is a factor
$3$ because there are at most two terms we have neglected, each of which is at most equal in magnitude to the dominant one in the zone.
{It follows that the maximum error in $T(R)$ is $3^{1/4}-1\approx32\pc$.
This is the same magnitude of error as in of~\citet{1993ASPC...36..371P} and~\citet{Haiman2009}, and is characteristic of matched-power law solutions.

We show in} Appendix~\ref{sec:solution-accuracy-checks} that the maximum error in $T(R)$ our \IRAS{} model is $20\pc$.
Given the uncertainties in the parameters used to construct the disc, such as
mass and viscosity which are known only to order-of-magnitude at best,
the speed advantage of our solution method compared to more complicated,
albeit more precise, methods makes it ideal for modelling the large
binary-star parameter space.

\subsubsection{Mass scaling}

\label{subsec:Mass-scaling}The term $\Sigma\left(R\right)$ is related
to the angular momentum flux through the disc $\mathcal{F}$ by,
\begin{alignat}{1}
\Sigma(R)  =\frac{\mathcal{F}}{3\pi h\nu}\,,\label{eq:SigmaR-from-flux}
\end{alignat}
where,
\begin{equation}
h=h(R)  =\sqrt{G\left(M_{1}+M_{2}\right)R}\,,\label{eq:h-specific-angular-momentum}
\end{equation}
is the specific angular momentum at radius $R$ and $\mathcal{F}$ is defined in \eqq{}~\ref{eq:angmomflux}. By combining this
with \eqq{}~\ref{eq:nu(R)}, we have,
\begin{equation}
\Sigma(R)  =\left(\frac{\mathcal{F}\mu}{3\pi\adisc\Gamma k_{\mathrm{B}}}\right)T^{-1}R^{-2}=\Sigma_{0}T^{-1}R^{-2}\,,\label{eq:SigmaR-power-law}
\end{equation}
which defines $\Sigma_{0}$,
\begin{equation}
\Sigma_{0}  =\Sigma(R)\,TR^{2}=\frac{\mathcal{F}\mu}{3\pi\adisc\Gamma k_{\mathrm{B}}}\,.\label{eq:Sigma0}
\end{equation}
The mass of the disc is set by choosing $\Sigma_{0}$ appropriately
as described in \secref{}~\ref{subsec:Constructing-the-disc}.

\subsubsection{Temperature power law $T\left(R\right)$}

\label{subsec:Temperature-power-law}By combining Eqs.~\ref{eq:T4eq}
and \ref{eq:SigmaR-power-law}, we select the temperature as a function
of radius according to the dominant term in \eqq{}~\ref{eq:T4eq}, 
\begin{equation}
T\left(R\right)  =\zeta R^{\eta}=  \max\left(AR^{-11/10},BR^{-3/7},CR^{-3/4}\right)\,.\label{eq:TR-power-law}
\end{equation}
The coefficients $A$, $B$, $C$, hence $\zeta$ and $\eta$, are given
in Table~\ref{tab:Power-law-summary}. 
\begin{table*}
  \caption{\label{tab:Power-law-summary}Temperature, $T$, power law prefactors,
$\zeta$, and slope exponents, $\eta$, where $T=\zeta R^{\eta}$
inside the disc and $R$ is the distance to the centre of the system.}
\begin{tabular}{cccc}
\hline 
Power law & Prefactor & Exponent & \makecell{Physical process}\tabularnewline
\hline 
 &  &  & \tabularnewline
$A$ & $A=a^{1/5}\Sigma_{0}^{2/5}=\left(\dfrac{27\kappa\adisc\Gamma\kb\sqrt{GM_{\star}}}{64\sigma\mu}\right)^{1/5}\Sigma_{0}^{2/5}$ & $-11/10$ & Viscous Heating\tabularnewline
$B$ & $B=b^{2/7}=\left(\dfrac{F_{\star}R_{\star}^{2}}{7\sigma}\sqrt{\dfrac{\Gamma\kb}{\mu GM_{\star}}}\right)^{2/7}$ & $-3/7$ & Irradiation (Bare Term)\tabularnewline
$C$ & $C=c^{1/4}=\left(\dfrac{2F_{\star}R_{\star}^{3}}{3\pi\sigma}\right)^{1/4}$ & $-3/4$ & Irradiation (Slope Correction) \tabularnewline
 &  &  & \tabularnewline
\hline 
\end{tabular}\end{table*}
We then write the temperature power law in zone $n$, which extends
from radii $R_{0,n}$ to $R_{1,n}$ (\eqq{}~\ref{eq:general-powerlaw}), with
exponent $a_{n}$ as,
\begin{equation}
T_{n}(R) =T_{0,n}\left(\frac{R}{R_{0,n}}\right)^{a_{n}}\,,\label{eq:Tlaw-full}
\end{equation}
where $T_{0,n}$ is the temperature at the inner edge and $T_{n}\left(R_{1,n}\right)$
is the temperature at the outer edge of the zone.

\subsubsection{General power laws}

\label{subsec:General-power-laws}Physical quantities throughout the
disc are expressed as power law functions of radius, $R$. Our discs
have $N$ zones labelled $n=0\dots N-1$ where $N\leq3$. A quantity
$A_{n}(R)$ in zone $n$ is represented by a power law in $R$, 
\begin{equation}
A_{n}(R)  =A_{0,n}\left(\frac{R}{R_{0,n}}\right)^{b_{n}}=A_{0,n}R_{0,n}^{-b_{n}}\,R^{b_{n}}\,,\label{eq:general-powerlaw}
\end{equation}
where $R_{0,n}$ and $R_{1,n}$ are the inner and outer radii of zone
$n$ such that in the zone $R_{0,n}\leq R\leq R_{1,n}$, $A_{0,n}=A_{n}\left(R_{0,n}\right)$
and $A_{1,n}=A_{n}\left(R_{1,n}\right)$ are the quantity at the inner
and outer radius of zone $n$, respectively, and $b_{n}$ is the exponent
of the power law in zone $n$. The power-law exponent can be written,
\begin{equation}
b_{n}  =ma_{n}+c\,,\label{eq:bn-from-an}
\end{equation}
where $m$ and $c$ are constants, and $a_{n}$ is the temperature
power-law exponent (Table~\ref{tab:Power-law-summary}). (Table~\ref{tab:Power-law-summary}).
Each physical quantity of interest can thus be defined by the two constants $m$ and $c$.

The power laws are continuous across zone boundaries. Crossing radii,
$R_{x,ij}$, between zones $i$ and $j$ satisfy,
\begin{equation}
A_{0,i}\left(\frac{R_{x,ij}}{R_{0,i}}\right)^{b_{i}}  =A_{0,j}\left(\frac{R_{x,ij}}{R_{0,j}}\right)^{b_{j}}\,,
\end{equation}
hence
\begin{equation}
R_{x,ij}  =\left(\frac{A_{0,i}}{A_{0,j}}\frac{R_{0,j}^{b_{j}}}{R_{0,i}^{b_{i}}}\right)^{\frac{1}{b_{j}-b_{i}}}\,.
\end{equation}
As an example, the surface density $\Sigma\left(R\right)$ is, from
\eqq{}~\ref{eq:SigmaR-power-law}, given by,
\begin{alignat}{1}
\Sigma(R) & =\Sigma_{0}\times T_{0,n}^{-1}\left(\frac{R}{R_{0,n}}\right)^{-a_{n}}\times R_{0,n}^{-2}\left(\frac{R}{R_{0,n}}\right)^{-2}\nonumber \\
 & =\Sigma_{0}T_{0,n}^{-1}R_{0,n}^{-2}\left(\frac{R}{R_{0,n}}\right)^{-a_{n}-2}\nonumber \\
 & =\Sigma\left(R_{0,n}\right)\left(\frac{R}{R_{0,n}}\right)^{-a_{n}-2},\label{eq:surface-density}
\end{alignat}
which is a general power law with $m=-1$ and $c=-2$ (\eqq{}~\ref{eq:bn-from-an}).
Further physical quantities are
given in Appendix~\ref{sec:Power-laws-appendix}.

\subsection{Integrated quantities}

\label{subsec:Integrated-quantities}Integrated properties of the
disc, such as its total mass, luminosity and angular momentum, are
constructed from our power-law solutions. The total integral, $I$,
over the disc is the sum of the partial integrals, $I_{n}$, in its
zones,
\begin{equation}
I =\sum_{n=0}^{N-1}I_{n}\,.
\end{equation}
We can write each integral using an integrand function, $i_{n}(R)$, 
which is also a power law in $R$,
\begin{equation}
i_{n}\left(R\right) =i_{0,n}\,\left(\frac{R}{R_{0,n}}\right)^{b_{n}}\,,
\end{equation}
hence over a single zone,
\begin{alignat}{1}
I_{n} & =\int_{R_{0,n}}^{R_{1,n}}i_{n}(R)\,\deriv R\\
 & =i_{0,n}\int_{R_{0,n}}^{R_{1,n}}\left(\frac{R}{R_{0,n}}\right)^{b_{n}}\,\deriv R\label{eq:In-with-limits}\\
 & =\frac{i_{0,n}}{1+b_{n}}R_{0,n}^{-b_{n}}\left(R_{1,n}^{1+b_{n}}-R_{0,n}^{1+b_{n}}\right)\,,
\end{alignat}
such that the total integral is,
\begin{equation}
I  =\sum_{n=0}^{N-1}  \frac{i_{0,n}}{1+b_{n}}R_{0,n}^{-b_{n}}\left(R_{1,n}^{1+b_{n}}-R_{0,n}^{1+b_{n}}\right) \,.\label{eq:general-integrand}
\end{equation}
In the $n$th zone only the constants $i_{0,n}$ and $b_{n}$ are
required to calculate the integrals $i_{n}$ and hence the total integral
$I$. In general we can then write,
\begin{equation}
b_{n} = ca_{n}+m\,,\label{eq:c-and-m}
\end{equation}
such that integral properties of the disc each correspond to a choice
of the constants $c$ and $m$.

\subsubsection{Total disc mass $\protect\Mdisc$}

\label{subsec:Total-disc-mass}The total mass of the disc is,
\begin{alignat}{1}
\Mdisc & =\int_{R_{\mathrm{in}}}^{R_{\mathrm{out}}}\deriv M=\int_{R_{\mathrm{in}}}^{R_{\mathrm{out}}}2\pi R\Sigma(R)\,\deriv R\\
 & =\sum_{n=0}^{N-1}\int_{R_{0,n}}^{R_{1,n}}2\pi\,R_{0,n}\left(\frac{R}{R_{0,n}}\right)\,\Sigma(R_{0,n})\left(\frac{R}{R_{0}}\right)^{-2-a_{n}}\deriv R\\
 & =\sum_{n=0}^{N-1}\left[2\pi R_{0}\Sigma\left(R_{0,n}\right)\right]\int_{R_{0,n}}^{R_{1,n}}\left(\frac{R}{R_{0}}\right)^{-1-a_{n}}\deriv R\,,\label{eq:mass-integral}
\end{alignat}
The integrand is in the form of \eqq{}~\ref{eq:general-integrand} with,
\begin{equation}
b_{n} = -a_{n}-1 = ca_{n}+m\,,\label{eq:bncm}
\end{equation}
where, 
\begin{equation}
c  =  m = -1\,,\label{eq:cm}
\end{equation}
 and $i_{n}=2\pi R_{0,n}\Sigma\left(R_{0,n}\right)$ (\eqq{}~\ref{eq:surface-density}).

\subsubsection{Further integrals}

\label{subsec:Further-integrals}Similarly to above, further global
disc properties can be written in this integral form, including the
angular momentum, angular momentum flux, luminosity, moment of inertia,
and the potential and kinetic energies of the disc. Appendix~\ref{sec:Integrals-over-the-disc}
derives them in full while table~\ref{tab:Integrand-functions} lists
the parameters $c$ and $m$ to be used in Eqs.~\ref{eq:general-integrand} and \ref{eq:c-and-m}.
Quantities such as the half-mass and half-angular momentum radii are
found by using the above method with bisection on a variable upper
limit in the integral. 
\begin{table*}
\caption{\label{tab:Integrand-functions}Integrand parameters which should
be applied in \eqq{}~\ref{eq:general-integrand} to calculate the
integrated properties of our circumbinary discs as listed in the first
column. The prefactors depend on properties of the inner binary and
the temperature $T\left(R_{0,n}\right)$, surface mass density $\Sigma\left(R_{0,n}\right)$
and radii $R_{0,n}$ at the inner edge of the $n$th zone in the disc.}
\begin{centering}
\def\arraystretch{1.4}%
\begin{tabular}{ccccc}
\hline 
Integral & Symbol & $c$ & $m$ & Prefactor $i_{0,n}$\tabularnewline
\hline 
Mass & $\Mdisc$ & $-1$ & $-1$ & $2\pi R_{0,n}\Sigma\left(R_{0,n}\right)$\tabularnewline
Angular momentum & $\Jdisc$ & $-\frac{1}{2}$ & $-1$ & $2\pi\sqrt{GM_{\mathrm{b}}}R_{0,n}^{3/2}\Sigma\left(R_{0,n}\right)$\tabularnewline
Angular momentum flux & $\mathcal{F}$ & $-6$ & $-1$ & $\ftid\pi a^{4}q^{2}GM_{\mathrm{b}}\,\Sigma\left(R_{0,n}\right)R_{0}^{-4}$\tabularnewline
Luminosity & $\Ldisc$ & $1$ & $4$ & $4\pi\sigma T_{0,n}^{4}R_{0,n}$\tabularnewline
Moment of inertia & $I_{\mathrm{disc}}$ & $1$ & $-4$ & $2\pi R_{0}^{3}\Sigma\left(R_{0,n}\right)$\tabularnewline
Gravitational potential energy & $E_{\mathrm{disc,grav}}$ & $-2$ & $-1$ & $\pi GM_{\mathrm{b}}\Sigma\left(R_{0,n}\right)$\tabularnewline
Kinetic energy & $E_{\mathrm{disc,kin}}$ & $-2$ & $-1$ & $-2\pi GM_{\mathrm{b}}\Sigma\left(R_{0,n}\right)$\tabularnewline
\hline 
\end{tabular}
\par\end{centering}
\end{table*}

\subsection{Constructing the disc}

\label{subsec:Constructing-the-disc}We have three parameters in our
circumbinary disc model: the surface density scaling factor $\sigmaz$,
the inner radius $\rin$ and the outer radius $\rout$. We have three
constraints: the disc mass $\mdisc$, the disc angular momentum $\jdisc$
and the disc angular momentum flux $\mathcal{F}$ which is related to
the resonant torque exerted by the inner binary on the disc (\secref{}~\ref{subsec:Binary-interaction}
and Appendix~\ref{sec:Integrals-over-the-disc}). We employ the \emph{GNU
Scientific Library} $\code{multiroot}$ solvers to find the appropriate
$\sigmaz$, $\rin$ and $\rout$ given $\mdisc$, $\jdisc$ and $\mathcal{F}$
to typically one part in $10^{6}$. On the rare occasion that this
method fails, we revert to a slower bisection method and then resume
with the $\code{multiroot}$ solvers. 

\section{Disc\textendash binary interaction}

\label{sec:Disc=002013binary-interaction}

A key aspect of our model is its ability to describe the changes in a circumbinary disc and its inner binary as time progresses. We combine a rapid stellar population code with our disc model to achieve this with the immediate goal of modelling the disc around \IRAS{}. 

\subsection{Stellar evolution and mass loss}

\label{subsec:Stellar-evolution}We evolve binary systems using $\binaryc$,
a rapid stellar evolution and nucleosynthesis code originally based
$\BSE$ of \citet{2002MNRAS_329_897H} with modifications as described
in \citet{2004MNRAS.350..407I,2006A&A...460..565I,2009A&A...508.1359I,2018MNRAS.473.2984I}.
$\binaryc$ calculates the stellar luminosity, radius, temperature,
core mass and wind-loss rates of stars in the binary as a function
of time and models mass transfer by winds and Roche-lobe overflow.
Systems which undergo unstable mass transfer enter a common-envelope
phase, which is described by the energy-balance prescription of \citet[their \secref{}~2.7.1]{2002MNRAS_329_897H}.
The parameter $\ace$ is the fraction of the orbital energy that is
available to eject the common envelope. We set $\ace=1$ by default.
The parameter $\lce$
describes the binding energy of the envelope as a function of its
mass and radius \citep{2002MNRAS_329_897H}, and is calculated based
on the models of \citet{2000A&A...360.1043D}. The envelopes of TPAGB
stars are poorly bound compared to other stages of stellar evolution,
so $\lce$ is often large, sometimes as much as $100$, and hence
the envelope is easily ejected and merging systems \textendash{} when
ejection fails \textendash{} are rare. A fraction $f_{\mathrm{ion}}$,
by default $10\pc$, of the envelope's recombination energy is assumed
to aid its ejection resulting in efficient envelope ejection from
TPAGB stars with little orbital shrinkage.

Our common-envelope prescription differs from \citet{2002MNRAS_329_897H}
in that we do not assume the post-common envelope object is a fully-stripped
white dwarf with no hydrogen envelope. Instead, we leave enough mass
in the post-AGB envelope that it $95\,\pc$ fills its Roche-lobe.
Typically this mass is about $10^{-2}\mathrm{\,M_{\odot}}$. Hydrogen
burning and mass loss erode the hydrogen envelope as the star moves
blueward in the Hertzsprung-Russell diagram. The star becomes a white
dwarf once its hydrogen envelope is gone, and it then evolves along
a canonical white dwarf cooling track. In our binaries there is the
possibility of mass accretion onto the stars from the inner edge of
the circumbinary disc. This slows blueward progress of the star
by replenishing the stellar hydrogen envelope (\secref{}~\ref{finalfates}).

Our stellar wind prescription follows that of \citet{2002MNRAS_329_897H}, where the \citet{1993ApJ...413..641V} wind is most important on the AGB and the \citet{1990A&A...231..134N} Wolf-Rayet wind activates during the hot post-AGB phase. We discuss the impact of this choice in Secs.~\ref{subsec:Post-(A)GB-and-binary-evolution}~and~\ref{finalfates}.

\subsection{Disc mass loss}

\label{subsec:Mass-loss}We treat mass changes to the disc as either
a fast phenomenon that changes the disc instantaneously, or as a slow
change that does not affect the disc structure, i.e.~\eqq{}~\ref{eq:calD}
is small compared to Eqs.~\ref{eq:calA}, \ref{eq:calB} and \ref{eq:calC}.
The mass and angular momentum of the circumbinary disc then evolve
according to,
\begin{alignat}{2}
\Mdotdisc & = & \dot{M}_{\mathrm{slow}}+\dot{M}_{\mathrm{fast}},\label{eq:mdot-disc}\\
\dot{J}_{\mathrm{disc}} & = & \dot{J}_{\mathrm{slow}}+\dot{J}_{\mathrm{fast}}\,,
\end{alignat}
with slow mass-loss and angular-momentum-loss rates given by,
\begin{equation}
\dot{M}_{\mathrm{slow}}  =  \dot{M}_{\mathrm{visc}}+\dot{M}_{\mathrm{wind}},\label{eq:mdot-slow}
\end{equation}
and 
\begin{equation}
\dot{J}_{\mathrm{slow}}  =  \dot{J}_{\mathrm{visc}}+\dot{J}_{\mathrm{wind}}+\dot{J}_{\mathrm{torque}}\,,\label{eq:jdot-slow}
\end{equation}
respectively, and their fast (i.e.~instantaneous) equivalents,
\begin{equation}
\dot{M}_{\mathrm{fast}}  =  \dot{M}_{\mathrm{inner}}+\Mdot{ram},\label{eq:mdot-strip}
\end{equation}
and
\begin{equation}
\dot{J}_{\mathrm{fast}} =  \dot{J}_{\mathrm{inner}}+\Jdot{ram}.\label{eq:jdot-strip}
\end{equation}
The terms in Eqs.~\ref{eq:mdot-disc} to \ref{eq:jdot-strip} are
described below.

\subsubsection{Slow mass changes, $\dot{M}_{\mathrm{slow}}$}

\label{subsec:Slow-mass-changes}The slow terms, $\dot{M}_{\mathrm{slow}}$
and $\dot{J}_{\mathrm{slow}}$, are applied to the disc every timestep.
Inflow onto the central binary from the inner edge of the disc is
at a viscous rate,
\begin{equation}
\Mdot{visc}  =  -3\pi\fvisc\nu\left(\rin\right)\Sigma\left(\rin\right)\,,\label{eq:mdot-disc-visc}
\end{equation}
where $\fvisc\approx1$ is a parameter. We show in \secref{}~\ref{subsec:Disc-evolution}
that the time-scale for the disc to evaporate by this method is an
order of magnitude longer than the disc lifetime and many orders of
magnitude longer than the binary orbit. Accreted gas takes a fraction of its angular
momentum with it onto the inner binary, i.e.,
\begin{equation}
  \dot{J}_{\mathrm{visc}}  =  \fviscJ h_{\mathrm{visc}}\dot{M}_{\mathrm{visc}}\,,\label{eq:Jdot-visc}
\end{equation}
with $h_{\mathrm{visc}}=h\left(\rin\right)$ and $\fviscJ\approx1$. 
\secref{}~\ref{subsec:Binary-interaction} describes accretion onto the
stars.

Radiation, particularly far-ultraviolet and X-ray, from the inner
binary penetrates the disc and launches a wind from its surface. The
rate of mass loss, $\dot{M}_{\mathrm{wind}}$, and associated rate
of angular momentum loss, $\dot{J}_{\mathrm{wind}}$, are calculated
by integrating \eqq{}~B5 of Appendix~B2 in \citet{2012MNRAS.422.1880O}
over the disc modulated by a parameter $\fX\approx1$. The X-ray
luminosity of the central stars, $L_{\mathrm{X}}$, is calculated
assuming they are black bodies and X-rays have energies in the range
$0.1-100\,\mathrm{keV}$.

The resonant torque, $\Jdot{torque}$, is described in \secref{}~\ref{subsec:Binary-interaction}.  We also adjust the angular
momentum flux, $\mathcal{F}$ to account for the lost material (the appropriate integral
is given in Appendix~\ref{subsec:Angular-momentum-flux-integral-correction}).

\subsubsection{Fast mass loss, $\dot{M}_{\mathrm{fast}}$}

\label{subsec:Fast-mass-loss}Processes that act on timescales typically shorter
than the disc viscous time-scale are grouped into a mass-loss rate,
$\dot{M}_{\mathrm{fast}}$, and associated angular momentum loss rate,
$\dot{J}_{\mathrm{fast}}$. 

The disc is instantaneously stripped at
the outer edge by interaction with the interstellar medium if the
disc is large enough. Given an interstellar medium pressure $P_{\mathrm{ISM}}=3000\,k_{\mathrm{B}}\,\mathrm{K}^{-1}$ \citep{2005ARA&A..43..337C}
we find the radius, $R$, at which $P(R)=P_{\mathrm{ISM}}$ and remove mass
and angular momentum outside this radius. In our models
of $\IRAS$, this radius lies outside the disc so ram-pressure stripping is never experienced.

Similarly, if the disc is small enough to be very close to the Roche
lobes, i.e.~$R_{\mathrm{in}}<a_{\mathrm{L2}}$, where $a_{\mathrm{L2}}$
is the second Lagrange-point radius, we strip material inside $a_{\mathrm{L2}}$
and remove its mass and angular momentum. Our \IRAS{} model contains sufficient angular momentum to avoid this.

\subsubsection{Coupled mass loss}

\label{subsec:Coupled-mass-loss}When mass is lost by photoevaporation
at either edge of the disc we provide the option, enabled by default,
to strip the disc to the radius at which the mass-loss time-scale equals
the local viscous time-scale. This is necessary because if the mass
loss is faster than viscous spreading, the disc cannot expand to quickly
replace the lost mass, which is one of our main assumptions in constructing
the disc. Typically in such cases the whole disc evaporates in a few
years, although for our model of \IRAS{} turning this option off makes negligible difference.

\subsection{Binary interaction}

\label{subsec:Binary-interaction}Mass is accreted onto stars $1$
and $2$ at rates,
\begin{equation}
\Mdot 1  = f_{q}\dot{M}_{\mathrm{visc}}\,,\label{eq:m1dot}
\end{equation}
and
\begin{equation}
\dot{M}_{2} =\left(1-f_{q}\right)\dot{M}_{\mathrm{visc}}\,,\label{eq:m2dot}
\end{equation}
respectively, where $f_{q}$ is set by either $f_{q}=0.5q$ \citep[our default choice]{2015MNRAS.452.3085Y},
$f_{q}=q/(1+q)$ \citep{2015MNRAS.451.3941G} or $f_{q}=0.5$, and
$q=M_{2}/M_{1}$. The mass carries specific angular momentum $h_{\mathrm{visc}}=\fviscJ h\left(\rin\right)$
(\eqq{}~\ref{eq:Jdot-visc}) onto the inner binary, where $\fviscJ=0.6$
by default based on the suggestion of \citet{2019ApJ...871...84M}
that we should use $\fviscJ=0.4-0.8$, {although this parameter may also exceed 1 according to their Fig.~3}. Note, however, that jet formation
may reduce accretion even further and we do not model this \citep[cf.][]{2019MNRAS.483.5020S}.
In any case, observed post-(A)GB binaries have orbital periods of
$100-1000\,\days$ with $J_{\mathrm{orb}}\approx10^{52}\,\junits$.
If the whole circumbinary disc is accreted without any loss of material,
$\text{\ensuremath{\Jorb}}$ changes by at most a fraction $\fJ\ll1$
with a correspondingly small change in orbital period and separation.
Accretion onto the stars, and associated spin up, is treated similarly
to Roche-lobe overflow in $\BSE$ and $\binaryc$ (\citealt{2002MNRAS_329_897H}
and \secref{}~\ref{subsec:Post-(A)GB-and-binary-evolution}).

The torque imparted on the disc by the binary, $\Jdot{torque}$, is
calculated by integrating the specific torque of \citet{2002ApJ...567L...9A}
over the disc assuming it is thin and larger than the binary orbit,
i.e.~$R>a$,
\begin{equation}
\Lambda(R)  =\frac{1}{2}\ftid a^{4}q^{2}GM_{\mathrm{binary}}\,R^{-5}\,,\label{eq:armitage-torque-1}
\end{equation}
where the multiplier $\ftid\approx10^{-3}$ is a parameter.
{This choice for $\ftid$ puts the inner edge of the disc at about $2a$, but $\rin$ can be increased by increasing $\ftid$ should the inner edge be at larger radius, e.g.~at the inner dust edge \citep{2007A&A...474L..45D}.}

We pump the binary eccentricity, $e$, according to the prescription
of \citet{2013A&A...551A..50D}\textbf{ }for which, 
\begin{equation}
\frac{\dot{a}_{\mathrm{res}}}{a}  = -2\fres\frac{l}{m}\adisc a\Omega_{\mathrm{b}}R_{\mathrm{J1/2}}^{3/2}\frac{\Mdisc\left(M_{1}+M_{2}\right)}{M_{1}M_{2}}\,,\label{eq:adot-res}
\end{equation}
where $l=1$, $m=2$, $\Omega_{b}$ is the orbital angular frequency
of the binary and $\fres\approx1$ is a parameter. The half-angular
momentum radius, $R_{\mathrm{J1/2}}$, is the radius that contains
half the disc's angular momentum and is found by bisection (\secref{}~\ref{subsec:Further-integrals}).
The eccentricity pumping rate is then,
\begin{alignat}{1}
\dot{e}_{\mathrm{res}} & =2e\,g_{R}\,\frac{\dot{a}_{\mathrm{res}}}{a}\times\begin{cases}
-25\adisc^{-1} & e<0.1\sqrt{\adisc}\,,\\
0 & e>0.7\,,\\
\left(\frac{1-e^{2}}{e^{2}+0.01\adisc}\right)\left(\frac{l}{m}-\frac{1}{\sqrt{1-e^{2}}}\right) & \mathrm{otherwise}.
\end{cases}\label{eq:edot-res}
\end{alignat}
The factor, 
\begin{equation}
g_{R}  =\min\left(\frac{2a}{\rin},1\right)^{3}\,,
\end{equation}
reduces the rate of eccentricity pumping when $\rin$ moves outside
the $l=1$, $m=2$ resonance. \citet{2013A&A...551A..50D} assume
$\rin=2a$, i.e.~$g_{R}=1$.

Despite its simplicity, the above formalism agrees reasonably with
modern hydrodynamic simulations \citep{2017MNRAS.464.3343F}. We ignore
eccentricity pumping caused by accretion of material onto the binary
because \citet{2020A&A...642A.234O} find its effect, based on the
results of \citet{2020ApJ...889..114M}, to be negligible.

\subsection{Post-(A)GB and binary evolution}

\label{subsec:Post-(A)GB-and-binary-evolution}
The luminosity of the post-(A)GB star is set by its core mass, where we use the fitting formulae of \citet[their \secref{}~2.7.1]{2002MNRAS_329_897H} during the first giant branch and early-AGB, or \citet{2004MNRAS.350..407I} during the thermally-pulsing AGB.
The initial radius of the post-AGB star is determined by the post-common
envelope separation, which in turn depends on $\ace$
and $\lce$ (\secref{}~\ref{subsec:Stellar-evolution}).
The post-(A)GB stellar radius and luminosity are calculated as in
\citet[their \secref{}~6.3]{2000MNRAS.315..543H}. The binary has an eccentricity
$\epostCE$ just after the common envelope is ejected, with $\epostCE=10^{-5}$
by default. Hydrodynamic common-envelope models \citep[e.g.][]{2012ApJ...744...52P}
predict eccentricities of about $0.1$ but these simulations never
conclude full envelope ejection, so the true value is probably less because of subsequent tidal damping.

In the absence of a circumbinary disc a post-(A)GB star loses material
from its hydrogen-rich envelope in a stellar wind while hydrogen burning
eats into it from below. We use the $\binaryc$ default mass-loss prescription, which is based on that of \citet{2002MNRAS_329_897H}. On the AGB, the \citet{1993ApJ...413..641V} rate dominates, but as the star transitions to the post-AGB the \citet{1990A&A...231..134N} Wolf-Rayet rate takes over. This is typically a bit less than $10^{-7}\,\msun\mathrm{yr}^{-1}$ in our \IRAS{} simulations.
We discuss other mass-loss prescriptions in \secref{}~\ref{finalfates}.
Nuclear burning also consumes the envelope at about $10^{-7}\,\mathrm{M}_{\odot}\,\mathrm{yr}^{-1}$.
As the hydrogen envelope thins the star heats, eventually becomes a white dwarf, and subsequently cools and dims. 

When there is a circumbinary disc, mass flows from its inner edge
onto the post-(A)GB star (\eqq{}~\ref{eq:mdot-disc-visc}), rejuvenating
the hydrogen envelope and extending the post-(A)GB phase. If the post-(A)GB
star expands and overflows its Roche lobe as a result of this accretion,
we allow mass to transfer to its companion but do not repeat common-envelope
evolution. This seems reasonable because the envelope, even though
it is convective, has little enough mass to shrink in response to mass loss, but
we admit a more realistic treatment of such phases is currently lacking \citep[cf.][]{2018MNRAS.480.5176H}.

\subsection{Disc termination}

\label{subsec:Disc-termination}Our evolution algorithm must be fast
because we wish to evolve millions of binary stars in stellar population
simulations. We thus choose conditions under which disc evolution
is terminated to prevent excessive and unnecessary numerical computation
of discs which would not be observed. 

When the mass of the circumbinary disc falls below $\Mmin=10^{-6}\mathrm{\,M_{\odot}}$
or it dims below $\Lmin=10^{-4}\mathrm{\,L_{\odot}}$ we terminate
its evolution. Such discs are difficult to observe and, because of
their low mass, hardly affect subsequent binary evolution.

If the time-scale for disc evaporation by slow mass loss, $\mdisc/\dot{M}_{\mathrm{slow}}$
(\eqq{}~\ref{eq:mdot-slow}), is shorter than $1\,\yr$ we evaporate
the disc instantly, thus stopping its evolution. This happens when
the X-ray induced wind becomes strong as the post-(A)GB star heats:
typically the disc evaporates before $T_{\mathrm{eff}}\approx4\times10^{4}\,\mathrm{K}$. 

We also terminate disc evolution when, 
\begin{equation}
\fring = \frac{\rout}{\rin}-1<0.2\,.\label{eq:fRing}
\end{equation}
Such discs are narrow rings with which our solution method sometimes
struggles (\secref{}~\ref{subsec:Constructing-the-disc}), and observed post-AGB discs have $\fring\gg1$ so are not such objects.

\section{The circumbinary disc around IRAS 08544-4431}

\label{sec:Example_discs}ALMA CO molecular-line maps of the circumbinary
disc around IRAS 08544-4431 show it has a mass of about $0.02\mathrm{\,M_{\odot}}$
and surrounds a central binary with orbital period $508\,\days$, eccentricity $0.22$ and
total mass $1.8\mathrm{\,M_{\odot}}$ containing a $0.65\mathrm{\,M_{\odot}}$,
$7250\,\mathrm{K}$, F3, $1.2\times10^{4}\mathrm{\,L_{\odot}}$ post-AGB
star with $\FeH=-0.5$ ($Z\approx0.004$ assuming $Z_{\odot}=0.014$)
and a $1.15\mathrm{\,M_{\odot}}$ companion, assuming a distance of
$1100\,\mathrm{pc}$ \citep{2018A&A...614A..58B,2018A&A...616A.153K}. The orbital separation
is thus $a\approx325\mathrm{\,R_{\odot}}\approx1.5\,\AU$.

{The inner
  edge of the gas disc is not resolved but the inner edge of dust lies at $9\,\AU$ \citep{2007A&A...474L..45D}.
The} outer edge is at $4.0\times10^{16}\,\cm\approx5.7\times10^{5}\mathrm{\,R_{\odot}}\approx2700\,\AU$.
The disc mass is $0.6-2.0\times10^{-2}\mathrm{\,M_{\odot}}$. The
disc angular momentum is about $13\mathrm{\,M_{\odot}}\,\km\,\mathrm{s}^{-1}\,\AU\approx3.9\times10^{52}\,\junits$
compared to the orbital angular momentum of about $20\mathrm{\,M_{\odot}}\,\km\,\mathrm{s}^{-1}\,\AU\approx6.1\times10^{52}\,\junits$.
The cubic dependence of angular momentum on the assumed distance introduces
considerable uncertainty, as discussed in detail by \citet{2018A&A...614A..58B},
and, if a distance of $550\,\mathrm{pc}$ is assumed, the orbital
angular momentum may be as small as $1.6\mathrm{\,M_{\odot}}\,\km\,\second^{-1}\,\AU=4.8\times10^{51}\,\junits$. 

\subsection{Progenitor binary system}

\label{subsec:Progenitor-system}The mass and luminosity of the post-AGB
star, $0.65\mathrm{\,M_{\odot}}$ and $1.2\times10^{4}\mathrm{\,L_{\odot}}$
respectively, are too massive and too bright to support the idea that
mass transfer started during the first ascent of the red giant branch.
$\IRAS$ is thus a true post-AGB system. If mass transfer starts on
the early-AGB (EAGB), a star with a $0.65\mathrm{\,M_{\odot}}$ core
has a total mass of $2.4\mathrm{\,M_{\odot}}$ and a radius of about
$200\mathrm{\,R_{\odot}}$. This is shorter than the current orbital
separation, $330\mathrm{\,R_{\odot}}$, so is impossible if we assume
common-envelope evolution only shortens the separation. 
The star is thus likely a post-TPAGB star.

The TPAGB star had a core of mass $0.65\mathrm{\,M_{\odot}}$ when
it lost its envelope, so with a metallicity $Z=0.004$ must have had
an initial mass of less than about $2.8\mathrm{\,M_{\odot}}$ (cf.~\figref~2
of \citealp{2004MNRAS.350..407I} based on the models of \citealp{Karakas2002}).
Its initial mass must exceed $1.2\mathrm{\,M_{\odot}}$ to become
a giant before its companion evolves from the main sequence. 

\begin{figure}
\begin{centering}
\includegraphics[width=1\columnwidth]{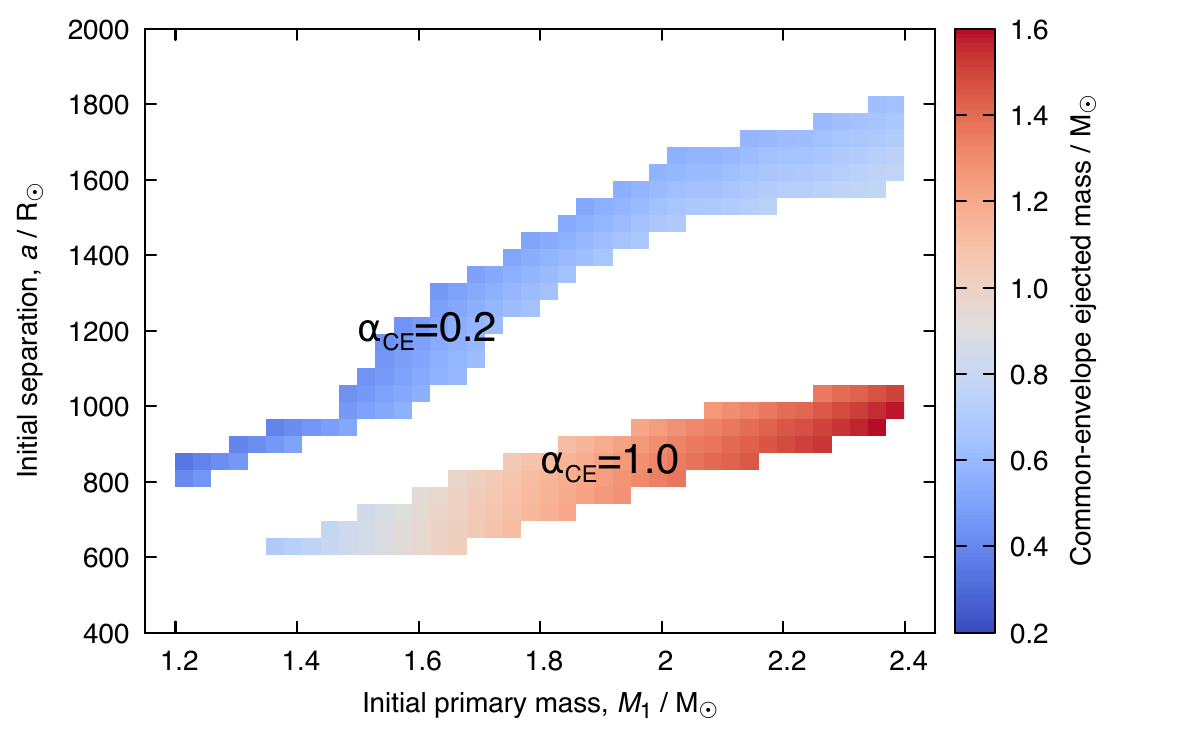}
\par\end{centering}
\caption{\label{fig:init-space-IRAS-mass}The initial separation and mass of
systems which, with our common-envelope prescription using $\protect\ace=0.2$
and $1.0$ and $f_{\mathrm{ion}}=0.1$, exit the common-envelope phase
with a post-AGB primary of mass $M_{1}=0.65\pm0.05\mathrm{\,M_{\odot}}$
and orbital separation $a=330\pm50\mathrm{\,R_{\odot}}$, assuming
an initial secondary and post-common envelope mass $M_{2}=1.15\mathrm{\,M_{\odot}}$.
The colour represents the mass of the ejected common envelope. These
systems are possible progenitors of $\protect\IRAS$ assuming about
$1\protect\pc$ of the common envelope mass forms a circumbinary disc. }
\end{figure}

In \figref~\ref{fig:init-space-IRAS-mass} we show the initial $a-M_{1}$
parameter space that corresponds to a post-AGB system with $M_{1}=0.65\pm0.05\mathrm{\,M_{\odot}}$
and $a=330\pm50\mathrm{\,R_{\odot}}$ with either $\ace=1.0$ or $0.2$ \citep{2011MNRAS.411.2277D,2012MNRAS.419..287D}.
When $\ace=0.2$, the energy available to eject the common envelope
is so limited that wind mass loss must remove much of the stellar
envelope before Roche-lobe overflow stars. Such systems typically
eject a common envelope of mass $0.2$ to $0.8\mathrm{\,M_{\odot}}$,
so to make a $0.02\mathrm{\,M_{\odot}}$ disc requires $\fM=0.025-0.10$.
When $\ace=1$ more energy is available to eject the common envelope
so wind mass loss is not as important and the common envelope mass
is $0.8-1.6\mathrm{\,M_{\odot}}$ corresponding to $\fM=0.0125-0.025$.
Without a good constraint on $\ace$ in AGB binaries we cannot say
which of the above scenarios is more likely, only that in both we
have sufficient mass to make the required circumbinary disc. 
\begin{figure}
\begin{centering}
\includegraphics[width=1\columnwidth]{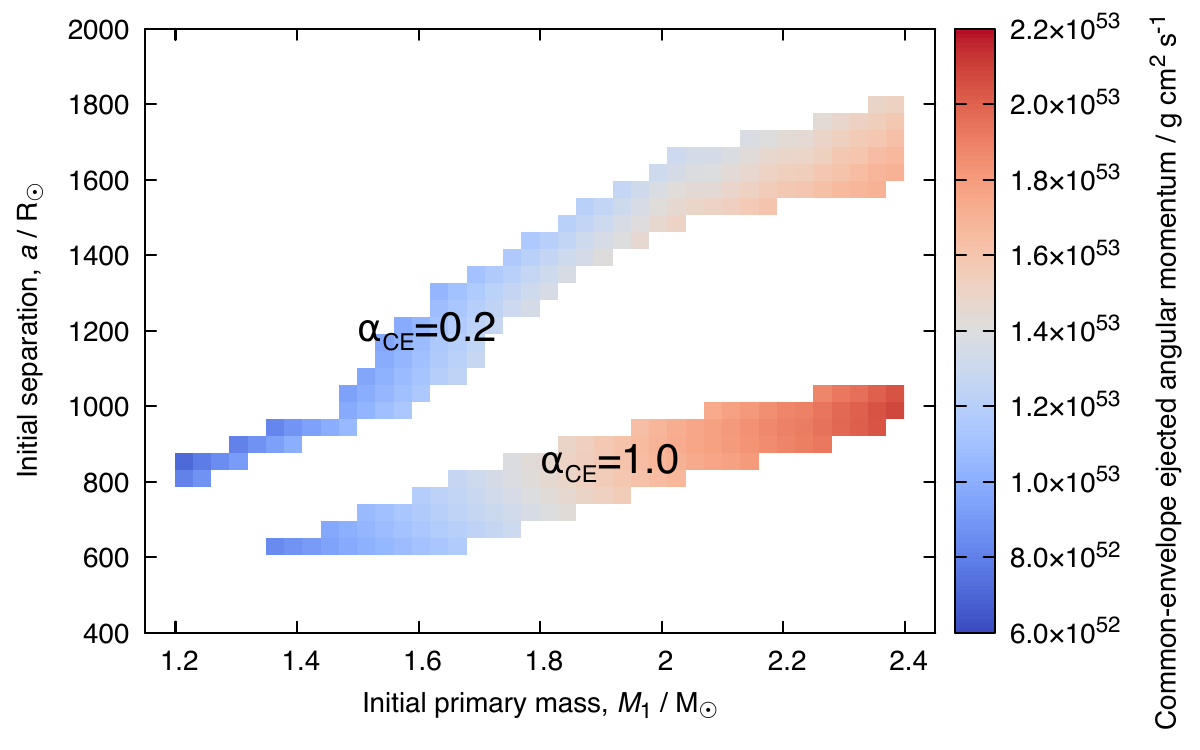}
\par\end{centering}
\caption{\label{fig:init-space-IRAS-angmom}As \figref~\ref{fig:init-space-IRAS-mass}
but showing angular momentum ejected during common-envelope evolution
rather than mass.}
\end{figure}

The angular momentum of the circumbinary disc around $\IRAS$ is
$0.48-6.1\times10^{52}\junits$. \figref~\ref{fig:init-space-IRAS-angmom}
shows that there is sufficient angular momentum in the common envelope
ejected from all our progenitor systems to make the disc. In both
our above example systems, $J_{\mathrm{orb}}\approx1.7\times10^{53}\,\jcgs$
so $2.8-36\pc$ of the orbital angular momentum goes into the circumbinary
disc.

In the following, we set initial stellar masses $M_{1}=2.00\mathrm{\,M_{\odot}}$
and $M_{2}=1.15\mathrm{\,M_{\odot}}$, initial separation $a_{\mathrm{init}}=900\mathrm{\,R_{\odot}}$,
metallicity $Z=0.004$ and adopt $\ace=1.0$. The post-AGB system
thus has $M_{1}=0.64\mathrm{\,M_{\odot}}$, $M_{2}=1.16\mathrm{\,M_{\odot}}$,
$L_{1}\approx0.95\times10^{4}\mathrm{\,L_{\odot}}$ and separation
$a=331\mathrm{\,R_{\odot}}$, which are similar to the properties
of the binary at the centre of $\IRAS$.

\subsection{Initial disc structure model}

\label{subsec:Initial-disc-structure}
\begin{figure*}
\begin{centering}
\includegraphics[viewport=45bp 65bp 648bp 725bp,scale=0.93]{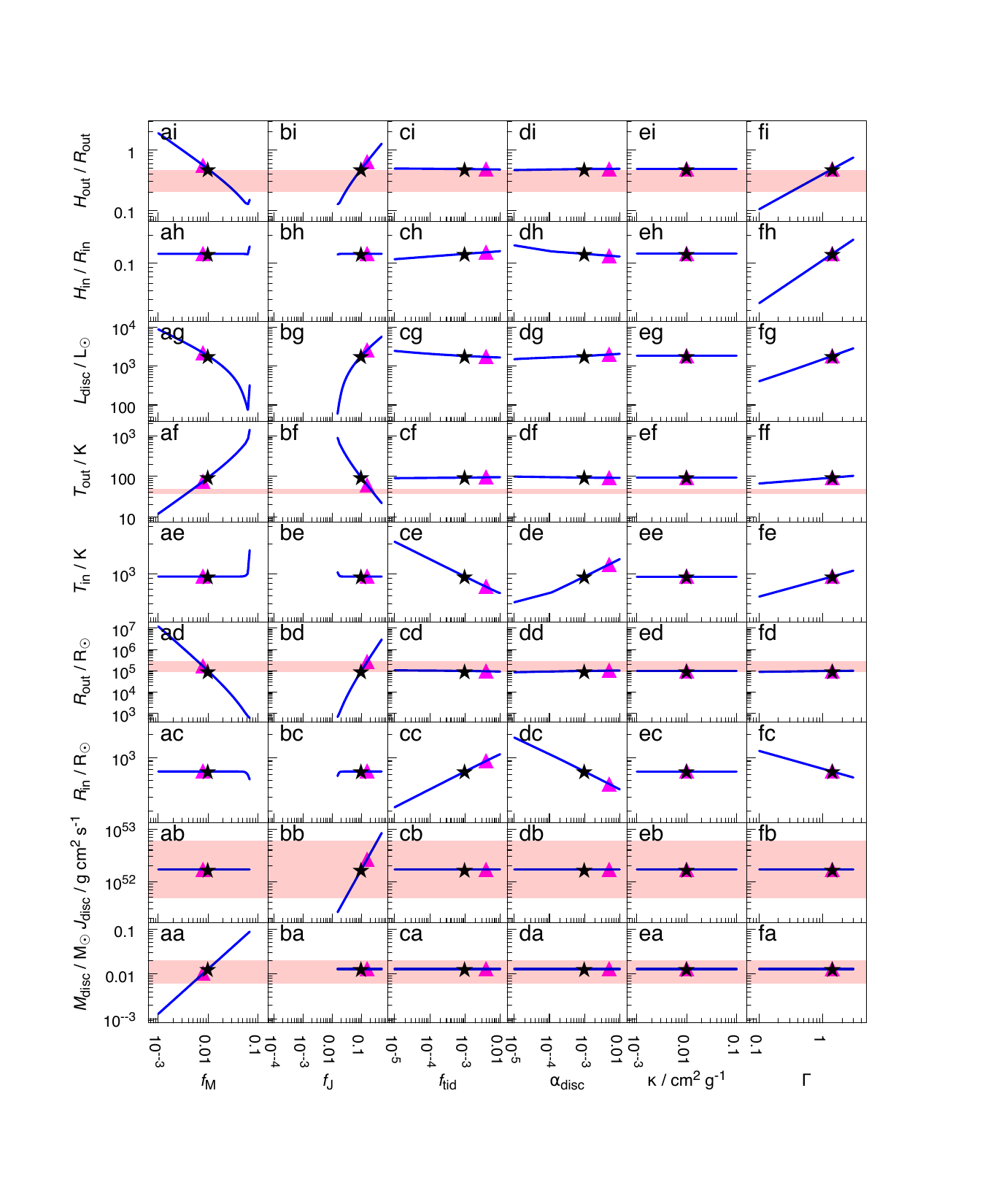} 
\par\end{centering}

\caption{\label{fig:disc_init_space}Sensitivity of {observable properties of our} circumbinary
disc {just after it forms, i.e.~when the common-envelope is ejected}, to {the main parameters} in our {disc} model. The blue
lines show how the observable (ordinate) {varies when each parameter
(abscissa) is changed in turn with all other parameters set to their defaults (Table~\ref{tab:disc-model-parameters}).}
The black stars are our initial
guess for the parameters (\secref{}~\ref{subsec:Initial-disc-structure})
while the magenta triangles are our best-fitting model parameters (\secref{}~\ref{subsec:Best-fitting-disc-parameters}).
The pink, shaded region represents the disc observed around
$\protect\IRAS$ where the range is mostly because of its uncertain
distance \citep{2018A&A...614A..58B}. $\protect\fM$ and $\protect\fJ$
are the fraction of the common-envelope mass and angular momentum
that are used to make the disc (Eqs.~\ref{eq:fM} and~\ref{eq:fJ}),
$\protect\ftid$ multiplies the torque that acts on the disc (\eqq{}~\ref{eq:armitage-torque-1}),
$\protect\adisc$ is the disc viscosity parameter (\eqq{}~\ref{eq:nu(R)}),
$\kappa$ is the disc opacity (\eqq{}~\ref{eq:calA}) and $\Gamma$
is the adiabatic index (\eqq{}~\ref{eq:H(R)}). $\protect\mdisc$ is
the disc mass, $\protect\jdisc$ is the disc angular momentum, $\protect\rin$
and $\protect\rout$ are the inner and outer radii of the disc, respectively,
$\protect\tin$ and $\protect\tout$ are the inner and outer edge
temperatures of the disc, respectively, and $\protect\hout/\protect\rout$
and $\protect\hin/\protect\rin$ are the ratios of the disc scale
height to its radius at its inner and outer edge, respectively.}
\end{figure*}

To initially model the disc around $\IRAS$
we set $\fM=0.01$ and $\fJ=0.10$, along with $\Gamma=1.4$, $\kappa=0.01\,\kappaunits$,
$\ftid=10^{-3}$ and $\adisc=10^{-3}$ (model parameters are listed
in table~\ref{tab:disc-model-parameters}). This results in a disc
of mass $0.013\mathrm{\,M_{\odot}}$ with $\jdisc=1.7\times10^{52}\,\junits=5.7\mathrm{\,M_{\odot}}\,\mathrm{km}\,\mathrm{s}^{-1}\,\AU$
. In \figref~\ref{fig:disc_init_space} we show how the initial properties
of the disc vary as the above parameters are varied in turn. Observables
correspond to the range of solutions found by \citet{2018A&A...614A..58B}
and we assume that the present-day $\IRAS$ differs little compared
to its birth state (\secref{}~\ref{subsec:Disc-evolution} justifies this).
\begin{table*}
\caption{\label{tab:disc-model-parameters}Parameters in our disc model, their
default values as used in our initial guess for the structure of the disc around \IRAS{}, the subsequent best fits to observed disc properties based on MCMC modelling (\secref{}~\ref{subsec:Best-fitting-disc-parameters}), and brief descriptions.}
\begin{tabular}{ccccc}
\hline 
Parameter & Default & Best fit & Definition & Description\tabularnewline
\hline 
$\fM$ & $0.010$ & $0.0079$ & \eqq{}~\ref{eq:fM} & Fraction of the common envelope mass that is put into the circumbinary
disc.\tabularnewline
$\fJ$ & $0.10$ & $0.16$ & \eqq{}~\ref{eq:fJ} & Fraction of the common envelope angular momentum that is put into
the circumbinary disc.\tabularnewline
$\kappa$ & $10^{-2}\,\kappaunits$ &  & \eqq{}~\ref{eq:calA} & Opacity in the circumbinary disc.\tabularnewline
$\Gamma$ & $1.4$ &  & \eqq{}~\ref{eq:H(R)} & Adiabatic index in the circumbinary disc, used to define the speed
of sound.\tabularnewline
$\adisc$ & $10^{-3}$ & $5.0\times10^{-3}$ & \eqq{}~\ref{eq:nu(R)} & Disc viscosity parameter.\tabularnewline
$\fvisc$ & $1$ &  & \eqq{}~\ref{eq:mdot-disc-visc} & \makecell{Multiplies the mass accretion rate from the inner edge of the circumbinary 
disc\\ onto the binary stars.}\tabularnewline
$\fviscJ$ & $0.6$ &  & \eqq{}~\ref{eq:Jdot-visc} & \makecell{Multiplies the angular momentum accretion rate from the inner edge
of the circumbinary disc\\ onto the binary stars.}\tabularnewline
$\ftid$ & $10^{-3}$ & $4.0\times10^{-3}$ & \eqq{}~\ref{eq:armitage-torque-1} & Multiplies the torque acting on the circumbinary disc.\tabularnewline
$\fres$ & $1$ &  & \eqq{}~\ref{eq:edot-res} & Multiplies the rate of eccentricity pumping of the binary star system.\tabularnewline
$\fX$ & $1$ &  & \secref{}~\ref{subsec:Slow-mass-changes} & Multiplies the X-ray flux incident on the disc.\tabularnewline
$\epostCE$ & $10^{-5}$ &  & \secref{}~\ref{subsec:Post-(A)GB-and-binary-evolution} & Eccentricity of the binary just after common envelope ejection.\tabularnewline
\hline 
\end{tabular}
\end{table*}

A narrow disc-mass range, $0.005\lesssim\fM\lesssim0.015$, matches
$\IRAS$ (\figref~\ref{fig:disc_init_space}~aa). To form a stable
disc, the angular momentum fraction $\fJ\gtrsim0.015$ such that $\jdisc\gtrsim2.5\times10^{51}\junits$
(\figref~\ref{fig:disc_init_space}~bb). Smaller discs are rings that
 fail the condition of \eqq{}~\ref{eq:fRing}.

Of the remaining parameters, $\kappa$ is irrelevant because the viscous
zone (\eqq{}~\ref{eq:a}) lies inside the binary orbit so is not in
the disc. The tidal parameter $\ftid$ and disc viscosity parameter
$\adisc$ affect the inner edge conditions, $\rin$ and $T_{\mathrm{in}}$,
which are not constrained by observations.
The inner radius lies at approximately twice the
orbital separation, $\rin\approx700\mathrm{\,R_{\odot}}\approx 2a \approx 3.3\,\AU$,
when $\ftid=10^{-3}$. When $\ftid\lesssim10^{-4}$ part of the disc
lies inside the binary orbit so is removed, reducing the mass of the
disc.  {\cite{2007A&A...474L..45D} find the inner edge of the dust to be at about $9\,\AU$, which is consistent with our model.    
  {Their disc model is in 2D with radiation transport, compared to our simpler but faster 1D model. Their 2D solutions include a puffed-up inner edge and shadowed region. Our disc inner-edge is always cooler than their 2D disc solution, $\tin \sim 800-900\,\mathrm{K}$ in our 1D model compared to their assumed $1500\,\mathrm{K}$ inner edge at the dust condensation temperature. The difference in $\tin$ between their model and ours, $\sim 600\mathrm{K}$, far exceeds the $\sim 200\,\mathrm{K}$ numerical uncertainties in our temperature solution. The differences most likely arise because we do not model the puffed-up edge which has $\hin /\rin =0.22$ in their 2D model compared to our simplified 1D disc which has $\hin/\rin\sim 0.11$.}
}

The initial outer edge properties, $\rout$, $\tout$ and $\hout/\rout$,
better match the observations if $\Gamma$ is reduced below its default
$1.4$, but this is difficult to justify physically. Our assumption that
the disc is Keplerian throughout, while $\IRAS$
is not quite Keplerian near its outer edge, could be responsible.
Our disc equations also assume $H/R\ll 1$
which is not strictly true \citep[e.g.][]{2021A&A...650L..13C}.
Increasing the angular momentum in our disc improves the match to $\rout$ (\figref~\ref{fig:disc_init_space}~bd) but worsens the matches to the temperature $\tout$ (\figref~\ref{fig:disc_init_space}~bf) and the ratio $\hout/\rout$ (\figref~\ref{fig:disc_init_space}~bi).

\subsection{Disc Evolution}
\label{disc-evolution}
\begin{figure}
\includegraphics[viewport=33bp 80bp 360bp 522bp,width=1\columnwidth]{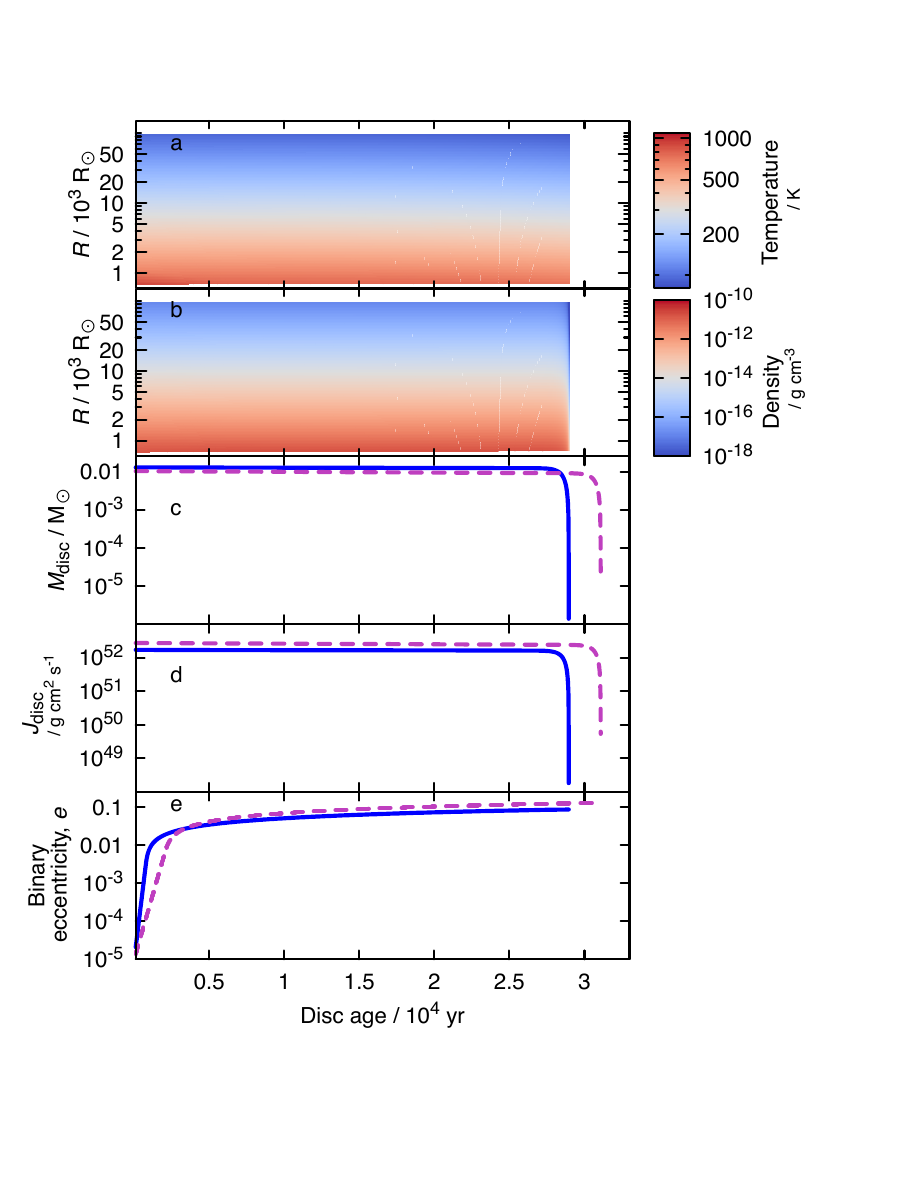}

\caption{\label{fig:Disc-T-rho-M_vs_t}Disc temperature (a) and density (b) as a function
of radius and time in our initial guess disc model, and disc mass (c), angular momentum (d) and binary eccentricity (e) as a function of time in our initial guess model (blue, solid lines) and best-fitting model (magenta, dashed lines) of $\IRAS$. In both models, the disc
hardly changes until the X-ray flux from the post-AGB star is sufficient
to evaporate it at about $3\times10^{4}\,\protect\yr$.  }
\end{figure}
\label{subsec:Disc-evolution}
The evolution of temperature, density
and total mass of our default model of $\IRAS$ is shown in \figref~\ref{fig:Disc-T-rho-M_vs_t}. 
Other than slow changes to its mass and angular momentum because of
flow through its inner edge and torque from the inner binary, respectively , our
disc is stable for most of its life. It loses mass through its inner
edge at a rate around $2.5\times10^{-8}\mathrm{\,M_{\odot}}\,\yr^{-1}$
(cf.~\eqq{}~\ref{eq:mdot-disc-visc}). If this were the only source
of mass loss, the disc would evaporate in about $4\times10^{5}\,\yr$.
Unfortunately for our disc, the post-AGB star's hydrogen-burning shell
eats through its hydrogen envelope more quickly, at a rate of about
$10^{-7}\mathrm{\,M_{\odot}}\,\yr^{-1}$, with a corresponding lifetime
of about $3\times10^{4}\,\yr$. As the envelope thins, the star heats and
its X-ray flux increases. X-ray induced mass loss evaporates
the disc after about $2.9\times10^{4}\,\yr$. By this time the
X-ray luminosity of the post-AGB star is around $0.01\mathrm{\,L_{\odot}}$
($3\times10^{31}\,\mathrm{erg}\,\second^{-1}$) which is sufficient
to evaporate similar protoplanetary discs \citep[cf.][]{2019MNRAS.483.3448M}. 

The final panel of \figref~\ref{fig:Disc-T-rho-M_vs_t} shows the binary eccentricity $e$ versus disc age.
Over its lifetime the disc pumps the binary eccentricity from $e=e_\mathrm{post-CE}=10^{-5}$ to nearly $0.1$.
The eccentricity of \IRAS{} is $0.22\pm0.02$ \citep{2018A&A...616A.153K}.
Increasing $e_\mathrm{post-CE}$ does not solve this discrepancy.
To reach $e=0.22$ requires a post-common envelope eccentricity $e_\mathrm{post-CE}=0.19$, double that seen in hydrodynamical simulations of common-envelope evolution \citep[e.g.][]{2012ApJ...744...52P}.
A factor $2.8$ increase to the rate of \eqq{}~\ref{eq:edot-res}, with $e_\mathrm{post-CE}=10^{-5}$, is sufficient to reach $e=0.22$.

\begin{figure*}%
\begin{raggedright}
\includegraphics[viewport=65bp 75bp 648bp 715bp,scale=0.95]{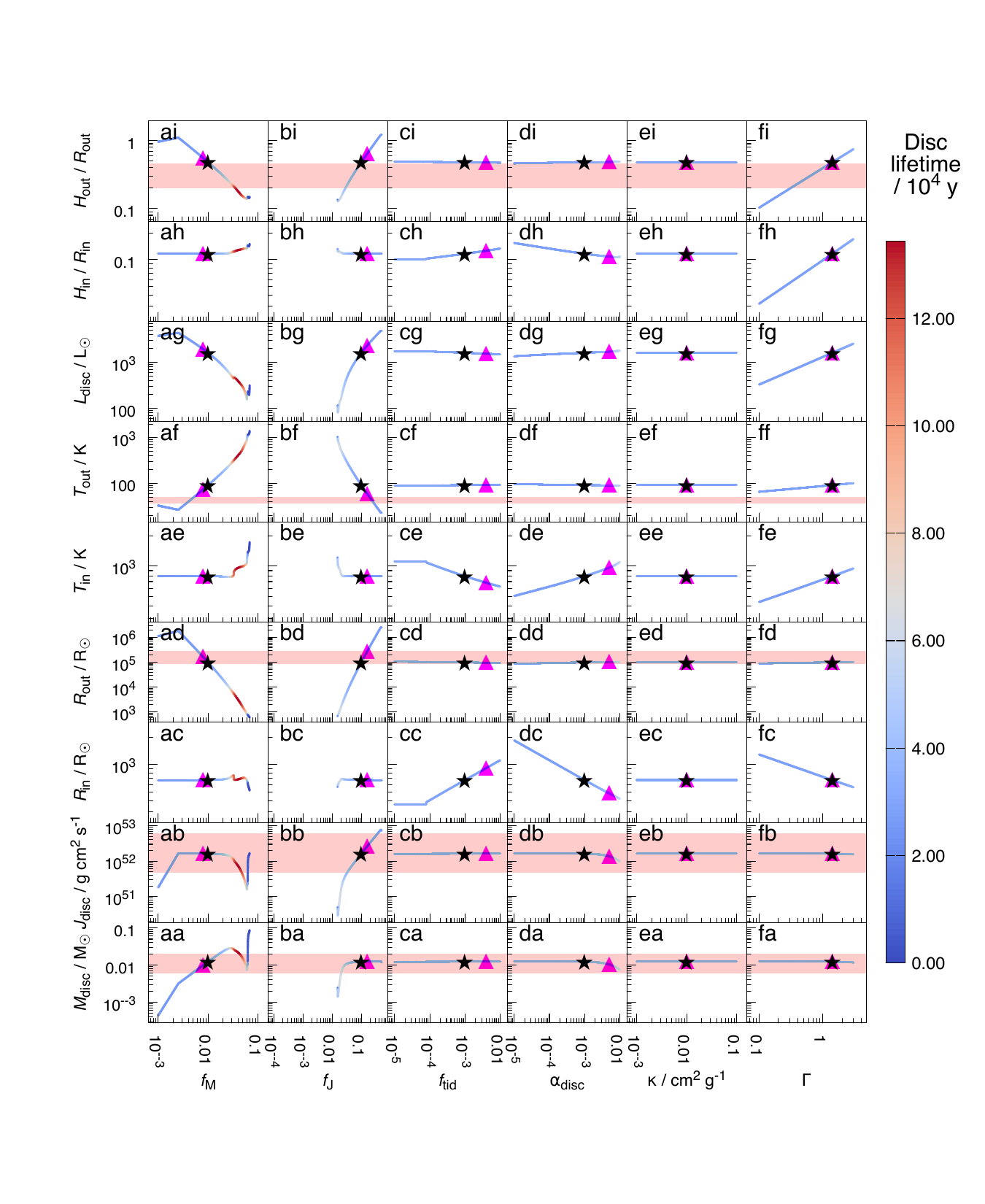}
\par\end{raggedright}
\caption{\label{fig:disc_mean_ev}As \figref~\ref{fig:disc_init_space} but instead
showing time-averaged properties of the our disc model. The line colour
represents the lifetime of the disc. The black star is our initial
guess for the parameters (\secref{}~\ref{subsec:Initial-disc-structure})
while the magenta triangle is our best-fitting model (\secref{}~\ref{subsec:Best-fitting-disc-parameters}).
{Most discs live for about $4\times10^4\,\mathrm{yr}$ because this is the time it takes the post-AGB star to heat sufficiently that the X-ray-induced disc-wind evaporates the disc. More massive discs, with $\fM\sim0.02-0.05$ have extended lifetimes because they can transfer more material onto the post-AGB star and keep it cool for longer (as discussed in Sec.~\ref{disc-evolution}). Discs can live for longer than shown here, $\gtrsim 1.2\times10^5\,\mathrm{yr}$, but only if they lose little or no mass which our default parameter choices, designed to match observed circumbinary discs, do not allow. The other parameters shown here change the disc lifetime very little.}
}
\end{figure*}

\figref~\ref{fig:disc_mean_ev} shows the time-averaged properties of
the disc and its lifetime as a function of its input parameters. As
described above, because the disc changes little over most of its lifetime, these properties are similar to the initial disc properties,
with a lifetime just short of $3\times10^4 \,\yr$. 

Massive discs, with $\fM\sim 0.03-0.05$ and $\fJ=0.1$, are exceptional. Because of their extra mass, the viscous inflow rate at the inner edge is similar to the burning rate of the hydrogen-burning shell of the post-AGB star, prolonging its life up to about $10^5\,\yr$. These discs are relatively compact, with outer radii shorter than $10^{4}\,\rsun$, so do not look like $\IRAS$ which has an outer radius in excess of $10^{5}\,\rsun$. However, if such discs are formed, their long lifetimes make them more likely to be observed and their stars' thick hydrogen envelopes keep their stars relatively cool ($\sim 3000\,\mathrm{K}$). We caution the reader: the sweet spot in $\fM$ that leads to long disc life corresponds to a particular choice of $\fJ$ and $\fvisc$. Choosing an even greater $\fM=0.06$ results in the disc being too compact to live very long -- it quickly forms a compact ring and is evaporated after about $10^3\,\yr$, while even more massive discs cannot form because they lack sufficient angular momentum. If, with $\fM=0.06$, we also set $\fJ=0.2$ the disc lives for  $1.8\times 10^{5}\,\yr$. If $\fvisc$ is smaller than the default, $1$, the mass of accreted hydrogen is reduced and the disc lives for less time as a result of the star reaching a temperature of $\sim 3\times10^{4}\mathrm{K}$ more quickly, at which point the disc is evaporated. The exception to this is if the disc runs out of mass, but this does not happen in any of our models with $\fM>10^{-4}$, $\fJ=0.1$ and $\fvisc<1$. These systems always evaporate their discs by X-ray-induced winds.

\subsection{Best-fitting disc parameters}

\label{subsec:Best-fitting-disc-parameters}We now attempt to fit
our disc parameters, $\fM$, $\fJ$, $\ftid$ and $\adisc$, to $\IRAS$ using our initial model as a first guess.
We match time-averaged disc properties from our model (\figref~\ref{fig:disc_mean_ev}) to
observed $\Mdisc$, $\Jdisc$, $\rout$, $\tout$ and $\hout/\rout$ \citep{2018A&A...614A..58B}.
We assume that each observable takes the mean value of the extremes
given by assuming distances of $550$ and $1100\,\mathrm{pc}$, we further assume
the associated error is Gaussian with the $1\sigma$ error set to half the range
of the observable. Thus $\mdisc=1.3\pm0.7\times10^{-2}\mathrm{\,M_{\odot}}$,
$\jdisc=3.29\pm2.81\times10^{52}\,\jcgs$, $\rout=1.3\pm0.7\times10^{16}\,\cm$,
$\tout=43\pm7\,\mathrm{K}$ and $\hout/\rout=0.33\pm0.13$. We further
can assume $\Mdotvisc=10^{-7}\pm0.5\times10^{-7}\mathrm{\,M_{\odot}}\,\yr^{-1}$
and $\rin/a=2.00\pm0.25$ as in previous works on circumbinary discs (e.g.~\citealt{2013A&A...551A..50D}). We employ $\emcee$ \citep{2019JOSS....4.1864F}
to perform a Markov-Chain Monte Carlo exploration of the parameter
space using 46 walkers with chain lengths of 2000, each of which has
its first 100 samples discarded. We assume flat priors in the ranges
$-7<\log_{10}\alpha<0$, $-5<\log_{10}\ftid<-1$, $-4<\log_{10}\fM<-1$
and $-4<\log_{10}\fJ<0$.

The resulting best-fit parameters are
$\log_{10}\adisc=-2.3_{-0.6}^{+1.4}$,
$\log_{10}\ftid=-2.4_{-0.6}^{+1.4}$,
$\log_{10}\fM=-2.1_{-0.3}^{+0.7}$
and $\log_{10}\fJ=-0.8_{-0.2}^{+0.6}$,
where the values quoted are
the means of the resulting distributions and uncertainties are $1\sigma$
(\figref~\ref{fig:gtc}). Without constraints on $\Mdotvisc$ and $\rin/a$
we cannot well constrain $\adisc$ or $\ftid$ (\figref~\ref{fig:gtc-no_Mvisc-no_Rina})
but find similar $\fM$ and $\fJ$: $\log_{10}\fM=-2.2_{-0.3}^{+1.2}$
and $\log_{10}\fJ=-0.8_{-0.3}^{+0.7}$.

\subsection{Sensitivity to evolutionary parameters}

\label{subsec:Sensitivity-to-evolutionary-parameters}In this section
we examine how our disc model reacts to reasonable changes in its
input parameters: $\adisc$, $\Gamma$, $\kappa$, $\fX$, $\fviscJ$,
$\fvisc$, $\fres$, $\ftid$, $\fJ$ and $\fM$. We use a Levenburg-Marquhart
algorithm \citep{1992nrca.book.....P} to fit functions
of the form,
\begin{equation}
\log_{10}\left(y\right) =\Phi+\Upsilon\log_{10}\left(x/x_{0}\right)+\Theta\left[\log_{10}\left(x/x_{0}\right)\right]^{2}\,,\label{eq:power-law-fit}
\end{equation}
where $y$ is a property of the disc that depends on parameter $x$
and $x_{0}$ is our default value of the parameter. $\Phi$, $\Upsilon$
and $\Theta$ are the constant parameters of each fit. We fit the time-averaged
inflow rate at the inner edge, $\overline{\dot{M}_\mathrm{visc}}$,  luminosity, $\overline{\Ldisc}$,
inner and outer radii, $\overline{\rin}$ and $\overline{\rout}$
respectively, scale height ratios at the inner and outer edge, $\overline{\hin/\rin}$
and $\overline{\hout/\rout}$ respectively, and the mean binary-star
eccentricity, $\overline{e}$. We also fit the maximum eccentricity
of the binary system, $e_{\mathrm{max}}$, the minimum Toomre parameter, $Q_\mathrm{min}$,  and
the disc lifetime, $\tdisc$. Appendix~\ref{sec:Fits} shows the
resulting fitting functions and lists the coefficients. In the following, we consider each
observable property of the disc, mostly in the form of power laws, $y\approx y_{0} x_{0}^{k}$, around our default model.

\begin{description}
\item[{\textbf{Disc lifetime}}] depends mostly on the mass in the disc. This is set by $\fM$ which governs the initial disc mass. Making the disc more compact by reducing $\fJ$ has a similar effect to increasing $\fM$. Secondly, mass loss from the inner edge of the disc, regulated through the viscosity parameter $\adisc$ and multiplier $\fvisc$, is also important. Together, the disc lifetime scales as  $2.9\times10^{4} \, (\adisc/0.001)^{0.13} \, (\fvisc)^{0.12} \, (\fJ/0.1)^{-0.21} \, (\fM/0.01)^{0.51} \,\yr$. The X-ray wind multiplier, $\fX$, has a relatively {small} impact on lifetime, {as long as $\fX$ is non-zero,} because the post-AGB X-ray flux increases extremely quickly as the star contracts and heats.

\item[\textbf{Disc size}] The inner radius, $\rin$, is a strong function of the tidal torque on the disc, through $\ftid$, and the gas properties $\adisc$ and $\Gamma$, such that $\overline{\rin} \approx 674 \, (\adisc/0.001)^{-0.21} \, (\Gamma/1.4)^{-0.24} \, (\ftid/0.001)^{0.22} \,\rsun$.
The outer radius depends primarily on the disc specific angular momentum through $\fJ$ and $\fM$, with $\overline{\rout} \approx 1.0\times 10^{5} \, (\fJ/0.1)^{2.3} \, (\fM/0.01)^{-2.2} \,\rsun$.

\item[\textbf{Disc luminosity}] is a function of disc size hence specific angular momentum and $\Gamma$, such that $\overline{\Ldisc} \approx 1640 \, (\Gamma/1.4)^{0.6} \, (\fJ/0.1)^{0.99} \, (\fM/0.01)^{-0.88} \,\lsun$.

\item[\textbf{Binary eccentricity}] is pumped to about $e=0.1$ if $\rin$ is inside the 2:1 Lindblad resonance, otherwise is less than $0.01$. The parameters that change $\rin$ are described above. We discuss eccentricity further in \secref{}~\ref{subsec:Discussion}.

\item[\textbf{Disc scale height ratios}] are particularly sensitive to $\Gamma$, and the outer edge ratio also depends sensitively on the disc specific angular momentum, such that $\overline{\hin / \rin} \approx 0.117 \, (\Gamma/1.4)^{0.49}$ and $\overline{\hout / \rout} \approx 0.483 \, (\Gamma/1.4)^{0.58} \, (\fJ/0.1)^{0.65} \, (\fM/0.01)^{-0.64}$.

\item[\textbf{Toomre's $Q$}] is defined such that $Q<1$ implies gravitational instability and possible planet formation \citep{1964ApJ...139.1217T}. We measure $Q_\mathrm{min}$, the minimum $Q$ at any point in the disc throughout its lifetime, and find this is sensitive to disc mass and specific angular momentum, as well as $\Gamma$, such that $Q_\mathrm{min} \approx 284 \, (\Gamma/1.4)^{0.61} \, (\fJ/0.1)^{0.92} \, (\fM/0.01)^{-1.9}$. All our simulated discs have $Q_\mathrm{min}\gg 1$, so are not expected to form planets by gravitational collapse, contradicting the assumption of \citet{2014A&A...563A..61S}.

\end{description}

\section{Discussion}
\label{subsec:Discussion}

Our disc model allows us to evolve many discs in a stellar population efficiently yet still reasonably well models the circumbinary disc around \IRAS{}. Below, we discuss some limitations, and strengths, of our model.

\subsection{Model limitations}

\label{subsec:Model-limitations}

  Our disc models assume constant opacity with a default $\kappa=0.01\,\kappaunits$ originating from gas with no dust contribution. Opacity is only important when energy transport is dominated by viscous heating (\eqq{}~\ref{eq:calA}), but in our disc around \IRAS{} this is never the case with $\kappa=0.01\,\kappaunits$. If we set $\kappa=10^{5}\,\kappaunits$ to force the entire disc to be one viscosity-dominated zone, our disc inner radius decreases from $\sim 650$ to $\sim 550\,\rsun$ and its luminosity increases from $1700\,\lsun$ to about $5000\,\lsun$. Other global properties such as the disc lifetime and its impact on the inner binary, e.g.~through mass accretion and eccentricity pumping, change little. A two-zone model containing an inner gas-dominated and outer dust-dominated structure is beyond the scope of our model, but based on the lack of significant changes to the binary interaction our model seems reasonable if the goal is to investigate feedback on the stars and their orbit.

  We also assume, for simplicity, that our disc is circular and aligned with the binary orbit. In our model, binaries eject their common envelopes with nearly-circular orbits -- we assume $\epostCE =10^{-5}$ -- and we perhaps naively neglect disc eccentricity.
  \citet{2008MNRAS.391..815P} suggest that discs should be roughly circular even in moderately eccentric binaries, while \citet{2020ApJ...905..106M} show that discs become eccentric even if the binary is circular because of eccentric mode trapping, and .
  
Many works, {e.g.~\cite{2015ApJ...800...96L}, \citet{10.1093/mnras/stv1450}, \cite{2015MNRAS.448.3472N} and \cite{2019ApJ...880L..18F}}, show that misinclination effects are complicated and highly non-linear, {particularly if a third star is present \citep{2022ApJ...927L..26M}}. Merging the results of such simulations into our model is beyond its scope at present, but is certainly worth future investigation.

 Our disc viscosity, modelled through $\adisc$, is assumed constant throughout the disc, but in reality depends on ionization fraction and hence the temperature distribution in the disc. Many features of the disc are sensitive to $\adisc$.  Relevant to the inner binary evolution is the increased efficiency of eccentricity pumping, and mass and angular momentum transfer from the inner edge to the stars. The central binary in our default model with $\adisc=10^{-3}$ suffers net loss of angular momentum from the binary to the disc because mass is accreted at about $2\times10^{-8}\,\msun\,\mathrm{yr}^{-1}$ which is insufficient to balance the torque imparted on the disc. With $\adisc=10^{-2}$ the situation is reversed and the inner binary gains angular momentum by accretion of mass at a rate of $1-3 \times10^{-7}\,\msun\,\mathrm{yr}^{-1}$. The accreted material keeps the central post-AGB star cool, thus limiting the wind from the disc and doubling its age to nearly $6\times10^4\,\mathrm{yr}$ (\secref{finalfates}). While we can, in theory, use the mass accretion rate in a comparison with observations to constrain $\adisc$, as in \figref~\ref{fig:gtc}, other observable disc properties such as the outer radius or outer temperature are not greatly affected by a change in $\adisc$. The post-AGB lifetime is also affected by accretion, so number counts of post-AGB binaries could also constrain $\adisc$ but, given the many biases and uncertainties in calculating such quantities, e.g.~assumed star-formation histories and selection functions, this will be a significant challenge. \cal{a}

 Our disc energy balance equation assumes all energy impacting the disc is absorbed and re-emitted, but perhaps we should assume a more realistic albedo. For an albedo $\psi$ we can simply scale the incoming stellar flux, $F_\star$, by $1-\psi$. The effect on disc properties is small because $L\sim T^4$ hence {$\delta T / T\sim \delta L / (4L)$}. Albedo is measured to be \textasciitilde 30\% in the R band in 89 Her, seen pole on (\citealt{2014A&A...568A..12H}).
 With an albedo of $0.5$ applied to our model of \IRAS{} the lifetime is reduced by only $500\,\mathrm{yr}$ and the eccentricity reaches $0.073$ compared to $0.087$. Both are negligible changes. 
  
The work of \cite{2021A&A...650L..13C} confirms that \IRAS{} has a puffed-up inner edge, which is not taken into account in our model and probably accounts for our (gas) inner edge temperature, about $1000\,\mathrm{K}$, being cooler than the observed \emph{H}-band temperature of $1290\pm70\,\mathrm{K}$ (this difference is small enough to be swallowed by our model uncertainty, \secref~\ref{subsec:Solution-method}). They also found that $T \propto r^{-0.8}$. Our best-fitting model of \IRAS{} starts with two zones, with exponents $-3/7\approx-0.43$ and $-3/4$, then at about $7000\,\mathrm{yr}$ has one zone with exponent $-3/4$. This latter solution agrees well with the observed temperature gradient.

\cite{2021AJ....161..147B} recently estimated the distance estimate of \IRAS{} to $1.53\pm 0.12 \,\mathrm{kpc}$ which, combined with the inner-edge radius estimate of \cite{2021A&A...650L..13C} gives $\rin \approx 2000 \rsun$. This is outside our gas-disc inner edge, at around $600\rsun$ in our best betting model, although our dust disc temperature at $2000\rsun$ is still only $765\mathrm{K}$, presumably because our model lacks the inner dust ring modelled in \cite{2021A&A...650L..13C}. 

\subsection{The origin of circumbinary-disc systems}
\label{sec:the-origin-of-circumbinary-disc-systems}

 We assume the disc around \IRAS{} is made by the ejection of a common-envelope
  but the disc could form from stellar winds \citep{2016A&A...596A..92K} if they
  start in a wide binary that shrinks to the present orbit
  \citep[e.g.][]{2010A&A...523A..10I}, or a binary that has stable mass transfer,
  perhaps with $L_2$ mass loss. This does not affect our results except from a binary evolutionary
  point of view. Similar systems to \IRAS{} have a large eccentricity, often $e>0.6$ \citep{2018A&A...620A..85O}, favouring the wind mass transfer
  hypothesis as tides are expected to be very efficient in stars close to, or overflowing, their Roche lobes.
  Unfortunately, wind mass transfer, often thought to perhaps be the origin of the eccentricity, seems not be able to overcome the circularization effect of tides either \citep{2019A&A...629A.103S}. 

 We use the fits to thin-envelope radius and luminosity, hence
  effective temperatures, of \cite{2002MNRAS_329_897H}.
  \cite{2016A&A...588A..25M} and \cite{2020A&A...642A.234O} show that the
  thermal pulse cycle, neglected in our model which assumes an average luminosity
  of the post-AGB star, is important in determining both the radius and luminosity. Given that
our modelled post-AGB star is of low mass, it has a long thermal pulse
cycle ($\gtrsim10^{4}\,\yr$) so is unlikely to have undergone more than a few thermal pulses while surrounded by a circumbinary disc. However, in the future this is something that should be taken into account in our models.

Finally, the question of whether \IRAS{} forms a planetary nebula, or ever had one, is an interesting one given that planetary nebulae typically have short-period central binaries (\citealt{2019ibfe.book.....B}; \citealt{2021MNRAS.506.5223J}). Our circumbinary disc lives for about $30,000\,\mathrm{yr}$, and its central star is only hot enough, $T_\mathrm{eff}\gtrsim 3\times10^4\,\mathrm{K}$, to ionise a planetary nebula in the last $5000\,\mathrm{yr}$ before the disc is evaporated. Given a typical dynamical lifetime of a planetary nebula in the LMC is about $3.5\times10^4\,\mathrm{yr}$, it is difficult to rule out formation through the binary channel with a central object with a period as long as the $500\,\mathrm{d}$ of \IRAS. The disc mass remaining when the nebula forms is near to its initial $0.01\,\msun$ but this drops quickly as the central star heats up, so it seems unlikely that a system will concurrently host a planetary nebula and a circumbinary disc. The lifetime of such planetary nebulae should also be shorter than in equivalent-mass single stars, because the nebula will have been expanding for most of the $2.5\times10^4\,\mathrm{yr}$ the disc has kept the its post-AGB star cool by accretion.

\subsection{Final fate, stellar winds and planetary survival}
\label{finalfates}
\begin{figure}
  \begin{centering}
    \begin{tabular}{c}
      \includegraphics[width=1\columnwidth]{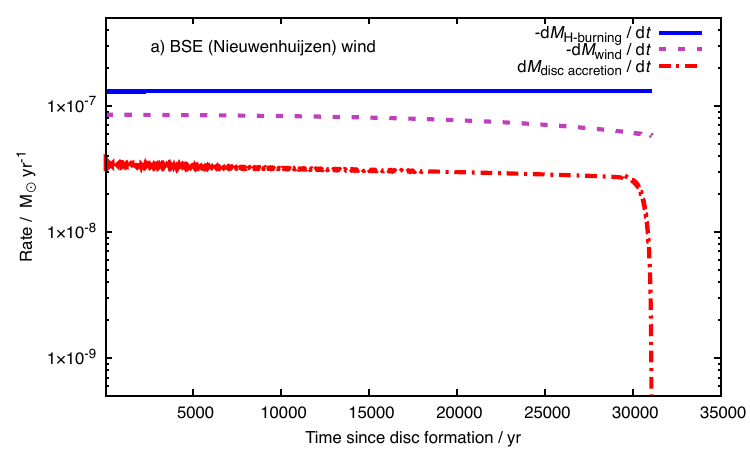}\tabularnewline
      \includegraphics[width=1\columnwidth]{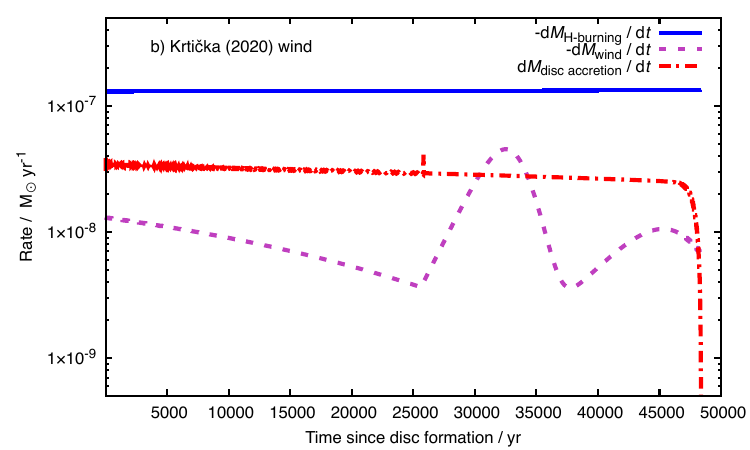}\tabularnewline
      \includegraphics[width=1\columnwidth]{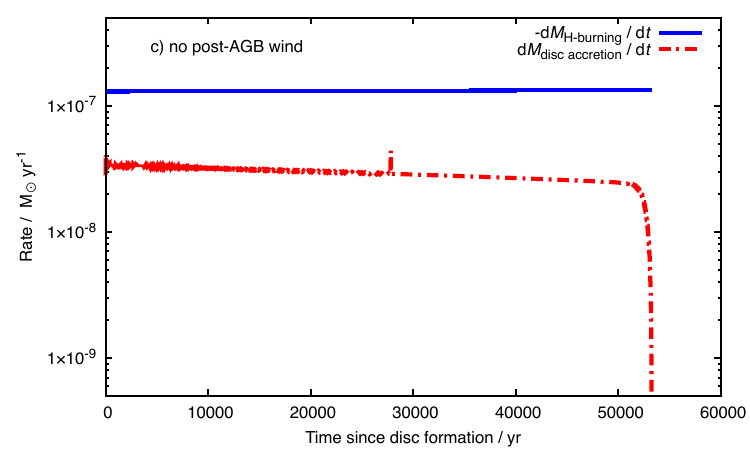}
    \end{tabular}
    \caption{\label{windcomp}{Comparison of the rate of post-AGB envelope destruction by hydrogen burning and wind mass loss, $dM_\mathrm{burning}/dt$ and $dM_\mathrm{wind}/dt$, and replenishment by accretion from the inner edge of the circumbinary disc, $dM_\mathrm{disc~accretion}/dt$,} in our best-fitting model of \IRAS{}. Panel (a) shows our default model that uses the \citet{2002MNRAS_329_897H} wind prescription in which the Wolf-Rayet wind of \citet{1990A&A...231..134N} dominates during the post-AGB phase, leading to disc evaporation after about $3\times10^{4}\,\mathrm{yr}$. Panel (b) uses the \citet{2020A&A...635A.173K} rate as an alternative which lengthens the post-AGB lifetime to about $5\times10^{4}\,\mathrm{yr}$ because the post-AGB star remains relatively cool for longer. Panel (c) has no post-AGB wind loss and is similar to (b).}
  \end{centering}
\end{figure}

Our post-common envelope, post-AGB star has a thin hydrogen envelope of mass $0.01\,\msun$. This is eaten from below at about $1.3\times 10^{-7}\,\msun\mathrm{yr}^{-1}$ by the hydrogen-burning shell that powers the stellar luminosity
and from above by stellar winds. In our standard wind model, based on \citet{2002MNRAS_329_897H}, the wind of \citet{1990A&A...231..134N} applies when the stellar luminosity exceeds $4\times10^{3}\,\lsun$ as in our \IRAS{} models which have $L_\mathrm{*}=7-9\times10^{3}\,\lsun$. This dominates the loss of the stellar envelope, as shown in \figref~\ref{windcomp}a, and is the main reason the post-AGB lifetime is limited to about $3\times10^4\,\mathrm{yr}$. As an alternative, we implemented the recently-determined rate of \cite[][our \figref~\ref{windcomp}b]{2020A&A...635A.173K}. This new rate is slow enough that it has little impact on the disc lifetime which extends to about $4.8\times10^{4}\,\mathrm{yr}$, similar to the $5.3\times10^{4}\,\mathrm{yr}$ lifetime with no post-AGB mass loss (\figref~\ref{windcomp}c). Also, with the \cite{2020A&A...635A.173K} rate the system reaches an eccentricity $e=0.27$ which is closer to that observed, $e=0.22$, in \IRAS{} than our default model.

\begin{figure}
  \begin{centering}
      \includegraphics[width=1\columnwidth]{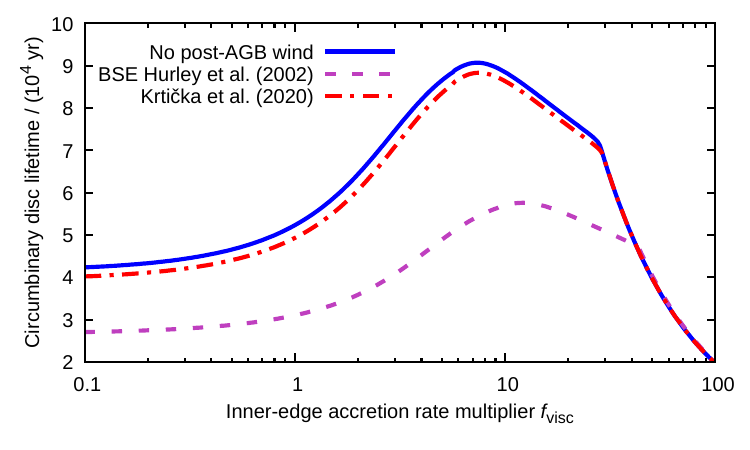}
      \caption{\label{fig:lifetimecomp}Circumbinary-disc lifetimes in our post-AGB models of \IRAS{} with no post-AGB wind (blue, solid line), the BSE wind of \citet[][red, dashed line]{2002MNRAS_329_897H} and the wind of \citet[][magenta, dot-dashed line]{2020A&A...635A.173K}, as a function of the viscous inner-edge accretion multiplier, $\fvisc$. At $\fvisc=0$ the lifetimes are equal to the post-AGB lifetime until $T_\mathrm{eff}\simeq 6\times10^{4}\,\mathrm{K}$.
If inner-edge accretion is 5 to 10 times faster than the local viscous rate at the inner edge of the disc, the post-AGB lifetime more than doubles. At faster accretion rates, the finite amount of mass in the disc limits its lifetime. Systems with $\fvisc\gtrsim 5$ have fast enough accretion that the post-AGB star resumes Roche-lobe overflow.} 
  \end{centering}
\end{figure}

The disc lifetime can be further extended by increasing the rate of mass inflow from the inner edge which then replenishes the post-AGB envelope, keeping the star cool and thus preventing disc evaporation (\figref~\ref{fig:lifetimecomp}). Using the \citet{2020A&A...635A.173K} wind rate with $\fvisc = 10$ (\eqq{}~\ref{eq:mdot-disc-visc}) indeed lengthens the life of the disc to about $8\times10^{4}\,\mathrm{yr}$. This is, however, not good for eccentricity pumping. The accretion onto the post-AGB star is so fast its envelope expands to fill its Roche lobe again, and in such circumstances tides prevent further eccentricity pumping which is limited to $e=0.16$ (cf.~\citealt{2020A&A...642A.234O}). Our treatment of Roche-lobe overflow assumes a circular binary so may be considered dubious in cases like this, but the amount of mass transferred is very small because the post-AGB envelope mass contains $\lesssim 0.01\,\msun$. Similar skepticism probably should probably be applied to our tidal formalism when the dissipating envelope is of such low mass. 

 Our use of a relatively simple X-ray induced wind prescription, designed for circumbinary discs around young-stellar objects, certainly could be criticised. However, the general form of such a wind -- in that it evaporates the disc as the star heats up -- is hard to deny. Once the disc mass is low enough, and accretion onto the central star is slower than the rate of hydrogen burning, the central star warms rapidly and mass loss is very quick (a few hundred years). Increasing or decreasing the X-ray wind rate by an order of magnitude makes negligible difference to the circumbinary disc lifetime for this reason. 

 The question of second-generation planet formation in binary systems is naturally fascinating, but with only around a dozen circumbinary planets observed such planets are apparently rare. {A fraction $\sim10\,\pc$ of post-AGB circumbinary discs may show signatures of such interactions because they lack infrared excess \citep{2022A&A...658A..36K}.}
Our model of \IRAS{} has a Toomre $Q$ of about $600$, suggesting the disc is unlikely to {suffer gravitational collapse} and subsequent planet formation. Planets around the original binary could, if they survive common-envelope evolution, accrete material from and carve gaps in the circumbinary disc, and perhaps explain the transition discs observed by \cite{2022A&A...658A..36K}.

Nonetheless, if an Earth-mass planet is at $1.1\,a_\mathrm{min}$ when the binary is born, where $a_\mathrm{min}\approx 2700\,\rsun$ is the minimum stable radius around our \IRAS{} binary \citep{2001MNRAS.321..398M}, its orbit expands to about $4700\,\rsun$ after common envelope ejection. It is thus in the circumbinary disc, near the inner edge, where accretion is possible and the disc is densest. Furthermore, the disc is cool enough, $T\sim 500\mathrm{K}$ at this radius, even in the \cite{2021A&A...650L..13C} model, to form dust grains and so perhaps accretion of these can continue even after the gas disc evaporates. Whether this process is efficient enough to be important is left to future work.

\subsection{Code performance} 

Using $\binaryc$ to model the binary system of $\IRAS$ to the end of the post-AGB phase takes, without disc evolution, about $0.5\,\mathrm{s}$ on an Intel i7-1185G7 $3.0\,\mathrm{GHz}$ CPU. Adding a circumbinary disc increases the run time to about $5.5\,\mathrm{s}$.  When simulating a population of binaries, only about $10\,\pc$ are expected to form circumbinary discs from common-envelope evolution, so the run time of a binary-population approximately doubles when circumbinary discs are included. {The estimate of $10\,\pc$ represents the fraction of binaries that are born in the appropriate period range that mass transfer occurs on the upper first-giant branch or AGB such that the common envelope is ejected rather than the stars merge. Assuming $M_{1}=1\msun$, $M_{2}=0.5\,\msun$ and $\ace=1$ and other physics as described as in \secref{}~\ref{subsec:Stellar-evolution}, these systems have periods $\log_{10}(P/\mathrm{d})\sim 2-3$, i.e. $\sim1\,\mathrm{dex}$. We usually model binary-star grids as a function of $0\lesssim \log_{10}P \lesssim 8$, so this is about $12.5\,\pc$ of these stars. Reducing $\ace$ to, say, $0.2$, changes the range to $\log_{10}(P/\mathrm{d})\sim 2.4-3$ or about $7.5\,\pc$.}

\section{Conclusions}

We present a fast evolutionary model of the circumbinary-disc system \IRAS{}. The properties of our modelled disc, such as its mass and angular momentum, its outer radius, scale height and temperature, and mass accretion rate on the inner binary, agree reasonably well with observations. Our modelled inner edge radius is consistent with observations. In about $3\times10^4\,\mathrm{yr}$ our circumbinary disc pumps the eccentricity of its inner binary to about $0.1$ when $\adisc=10^{-3}$, half that observed in \IRAS{}. A modest increase in $\adisc$ allows us to match both the eccentricity in \IRAS{} and a viscous-timescale dominated accretion rate of $10^{-7}\,\msun\mathrm{yr}^{-1}$, similar to that observed. Accretion onto the binary from the inner edge of the disc increases the duration of the post-AGB phase by only about $16\,\pc$ compared to an equivalent single star. This lifetime is, instead, dominated by nuclear burning through its hydrogen envelope and stellar wind mass loss. Increasing the rate of inner edge accretion by a factor $\sim10$ approximately doubles the lifetime of the post-AGB--circumbinary-disc system.   
We discuss both limitations and strengths of our model, the greatest of which is its speed. From birth to evaporation, our circumbinary-disc model takes about $5\,\mathrm{s}$ of computation time, making it suitable for binary-star population synthesis.

\section*{Acknowledgements}

\label{sec:Acknowledgements}

We thank the anonymous referee for their helpful comments.
RGI thanks the STFC for funding his Rutherford fellowship under grant
ST/L003910/1, helping him through the difficult Covid-19 period with grant ST/R000603/1, Churchill College, Cambridge for his fellowship and access to their library.
The Flatiron Institute is supported by the Simons Foundation.
ASJ thanks the Gordon and Betty Moore Foundation (Grant GBMF7392), the National Science Foundation (Grant No. NSF PHY-1748958) and the UK Marshall Commission for supporting this work.
This research has made use of NASA's Astrophysics
Data System Bibliographic Services. We thank Dave Green for his Cubehelix
colour scheme \citep{2011BASI...39..289G}. Many thanks go to, in
no particular order and for useful discussions and help with this project, to Hans Van Winckel, Clio Gielen, Jacques Kluska, Glenn-Michael Oomen,
Leen Decin, James Owen, Cathie Clarke, Richard Booth, Chris Tout,
Philipp Podsiadlowski, Jason Isaacs and Hagai Perets.

\section*{Data Availability}

\label{sec:Data-availability}

The $\binaryc$ code developed for this project is available at its website, \url{http://personal.ph.surrey.ac.uk/~ri0005/binary_c.html}, and its Gitlab pages \url{https://gitlab.eps.surrey.ac.uk/ri0005/binary_c} and \url{https://gitlab.com/binary_c}. Data used and shown in this paper is available on reasonable request to the lead author.

\bibliographystyle{mnras}
\bibliography{references}

\vfill{}

\appendix
\onecolumn

\section{Local disc properties}

\label{sec:Power-laws-appendix}Physical quantities can be calculated
at a radius $R$ given a disc temperature structure $T=T\left(R\right)$.
For example, the local disc scale height is,
\begin{equation}
H\left(R\right) =\frac{k_{\mathrm{B}}\Gamma}{\mu GM_{\star}}TR^{3}\,,\label{eq:scale-height}
\end{equation}
the sound speed is,
\begin{equation}
\cs\left(R\right)  =\sqrt{\frac{k_{\mathrm{B}}\Gamma T}{\mu}}\,,\label{eq:sound-speed}
\end{equation}
the density is,
\begin{equation}
\rho\left(R\right)  =\frac{\Sigma\left(R\right)}{H\left(R\right)}\,,\label{eq:density}
\end{equation}
the gas pressure is,
\begin{equation}
P\left(R\right)  =\frac{k_{\mathrm{B}}T\rho\left(R\right)}{\mu}\,,\label{eq:pressure}
\end{equation}
the kinematic viscosity (which requires \eqq{}~\ref{eq:Omega(R)}) is,
\begin{equation}
\nu\left(R\right)  =\frac{\adisc\cs^{2}\left(R\right)}{\Omega\left(R\right)}\,,\label{eq:kinematic-viscosity}
\end{equation}
the orbital velocity is,
\begin{equation}
v\left(R\right)  =R\Omega\left(R\right)\,,\label{eq:orbital-velocity}
\end{equation}
the local viscous time-scale is,
\begin{equation}
\tau_{\mathrm{visc}}\left(R\right)  =\frac{R^{2}}{5\nu\left(R\right)}\,,\label{eq:local-viscous-time-scale}
\end{equation}
and the \citet{1964ApJ...139.1217T} stability parameter, $Q$, is,
\begin{equation}
Q =\frac{\cs\left(R\right)\kappa}{\pi G\Sigma(R)}\,.\label{eq:ToomreQ}
\end{equation}

\section{Disc integrals}

\label{sec:Integrals-over-the-disc}In this section we provide derivations
of the parameters $c$ and $m$ to be used in \eqq{}~\ref{eq:c-and-m}
when calculating integral properties of the disc.

\subsection{Angular momentum $\protect\jdisc$}

The total angular momentum in the disc is, 
\begin{alignat}{1}
  \Jdisc & =\int h\,\deriv M\\
 & =\int_{R_{\mathrm{in}}}^{R_{\mathrm{out}}}2\pi hR\Sigma(R)\,\deriv R\\
 & =2\pi\sqrt{GM_{\mathrm{b}}}\int_{R_{\mathrm{in}}}^{R_{\mathrm{out}}}\Sigma(R)\,R^{3/2}\,\deriv R\\
 & =2\pi\sqrt{GM_{\mathrm{b}}}\sum_{n=0}^{N-1}\int_{R_{0,n}}^{R_{1,n}}\Sigma\left(R_{0,n}\right)\left(\frac{R}{R_{0,n}}\right)^{-2-a_{n}}\,R_{0,n}^{3/2}\left(\frac{R}{R_{0,n}}\right)^{3/2}\deriv R\\
 & =\sum_{n=0}^{N-1}\left[2\pi\sqrt{GM_{\mathrm{b}}}\,\Sigma\left(R_{0,n}\right)R_{0,n}^{3/2}\right]\int_{R_{0,n}}^{R_{1,n}}\left(\frac{R}{R_{0,n}}\right)^{-1/2-a_{n}}\deriv R\,,\label{eq:disc-angmom-integral}
\end{alignat}
i.e.~$b_{n}=-\frac{1}{2}-a_{n}$ (i.e.~$c=-1/2$ and $m=-1$ in
\eqq{}~\ref{eq:bncm}) with $i_{0,n}=2\pi\sqrt{GM_{\mathrm{b}}}\Sigma\left(R_{0,n}\right)R_{0,n}^{3/2}$.

\subsection{Angular momentum flux $\mathcal{F}$}

\label{subsec:Angular-momentum-flux-integral}The total angular momentum
flux through the disc is found by integrating the specific torque
throughout the disc,
\begin{equation}
  \mathcal{F}  =\int_{R_{\mathrm{in}}}^{R_{\mathrm{out}}}2\pi R\Sigma(R)\Lambda(R)\,\deriv R\,,\label{eq:angmomflux}
\end{equation}
where we assume the specific torque of \citet{2002ApJ...567L...9A},
\begin{alignat}{1}
\Lambda(R) & =\frac{\ftid q^{2}GM_{\mathrm{b}}}{2\Delta^{4}}\times\begin{cases}
R^{3} & R<a\,,\\
a^{4}R^{-1} & R>a\,,
\end{cases}\label{eq:armitage-torque}
\end{alignat}
where $q=M_{2}/M_{1}<1$, $\Delta=\max(H,\left|R-a\right|)$ and
$\ftid\approx10^{-2}$. Our disc is thin ($H/R\ll1$) and typically far from
the centre of mass ($R\gg a$) so $\Delta\approx R$ and the torque
is as in \eqq{}~\ref{eq:armitage-torque-1}.
The total angular momentum flux is then,
\begin{alignat}{1}
\mathcal{F} & =\pi\ftid a^{4}q^{2}GM_{\mathrm{b}}\int_{R_{\mathrm{in}}}^{R_{\mathrm{out}}}\Sigma(R)\times R\times R^{-5}\,\deriv R\\
 & =\pi\ftid a^{4}q^{2}GM_{\mathrm{b}}\int_{R_{\mathrm{in}}}^{R_{\mathrm{out}}}\Sigma\left(R_{0}\right)\left(\frac{R}{R_{0}}\right)^{-2-a_{n}}\times R_{0}^{-4}\left(\frac{R}{R_{0}}\right)^{-4}\deriv R\\
 & =\left[\pi\ftid a^{4}q^{2}GM_{\mathrm{b}}\Sigma\left(R_{0}\right)R_{0}^{-4}\right]\int_{R_{\mathrm{in}}}^{R_{\mathrm{out}}}\left(\frac{R}{R_{0}}\right)^{-6-a_{n}}\deriv R\,\label{eq:flux-integral}\\
 & =\pi\ftid a^{4}q^{2}GM_{\mathrm{b}}\sum_{n=0}^{N-1}\left[\Sigma\left(R_{0,n}\right)R_{0,n}^{-4}\right]\int_{R_{0,n}}^{R_{1,n}}\left(\frac{R}{R_{0,n}}\right)^{-6-a_{n}}\deriv R\,,
\end{alignat}
i.e.~$b_{n}=-6-a$ (i.e.~$c=-6$ and $m=-1$ in \eqq{}~\ref{eq:bncm})
and $i_{n}=\pi\ftid a^{4}q^{2}GM_{\mathrm{b}}\Sigma\left(R_{0,n}\right)R_{0,n}^{-4}$.

\subsection{Angular momentum flux correction}

\label{subsec:Angular-momentum-flux-integral-correction}

The flux in Appendix~\ref{subsec:Angular-momentum-flux-integral} assumes that the disc mass is fixed.
We expect to lose mass to both fast and slow processes, where fast means faster than the viscous time-scale and slow means on or slower than the viscous time-scale (\secref~\ref{subsec:Mass-loss}.
Fast processes do not change the angular momentum flux because the material they remove is simply gone, not replaced by the disc spreading.
Slow processes on the other hand do change the angular momentum flux, because the disc has time to adjust to fill in the mass that was lost.

To approximate the effect of these slow processes  we begin by noting that~\citep{2016ApJ...830....7R},
\begin{align}
    \frac{d\mathcal{F}}{dh} = \dot{M},
\end{align}
where $h = \sqrt{G M R}$ is the specific angular momentum of a Keplerian orbit and $\dot{M}$ is the amount of mass passing through the radius $R$ per unit time.

We assume a fixed $\dot{M}$ throughout the disc.
We expect this to be a good approximation for X-ray mass loss, which happens primarily near the inner edge of the disc, and it is likewise a good approximation for mass loss through the inner edge of the disc.
The correction to the angular momentum flux is then,
\begin{equation}
   \delta \mathcal{F} = \dot{M}_\mathrm{slow} (h - h_{\rm in})\,,
\end{equation}
where $\dot{M}_\mathrm{slow}$ is defined in \eqq{}~\ref{eq:mdot-slow}. Our disc model requires an \emph{average} angular momentum flux, so we integrate this with respect to $h$ in the disc to find,
\begin{equation}
  \delta \mathcal{F} \approx  \delta \mathcal{F}_{\rm avg} = \frac{1}{h_\mathrm{out} - h_\mathrm{in}}\int_{h_\mathrm{in}}^{h_\mathrm{out}} \delta \mathcal{F}(h)\, dh = \frac{1}{2} (h_\mathrm{out} - h_\mathrm{in}) \dot{M}_{\rm slow}\,.
\end{equation}

\subsection{Disc total luminosity}

\label{subsec:Disc-total-luminosity-integral}The disc luminosity
is,
\begin{alignat}{1}
\Ldisc & =2\int_{R_{\mathrm{in}}}^{R_{\mathrm{out}}}2\pi R\sigma T^{4}\,\deriv R\\
 & =\int_{R_{\mathrm{in}}}^{R_{\mathrm{out}}}4\pi R\sigma T_{0}^{4}\left(\frac{R}{R_{0}}\right)^{4a_{n}}\,\deriv R\\
 & =\sum_{n=0}^{N-1}\left(4\pi\sigma T_{0,n}^{4}R_{0,n}\right)\int_{R_{0,n}}^{R_{1,n}}\left(\frac{R}{R_{0}}\right)^{4a+1}\deriv R\,,
\end{alignat}
where the factor $2$ is because the disc has two sides. The integrand
is in the form of \eqq{}~\ref{eq:general-integrand} with $b_{n}=1+4a$
i.e.~$m=4$ and $c=1$ in \eqq{}~\ref{eq:bncm}, and $i_{n}=4\pi\sigma T_{0,n}^{4}R_{0,n}$.

\pagebreak{}

\section{Fits of disc properties to model parameters}

\label{sec:Fits}
\begin{figure}
  \centerfloat
  \begin{centering}
\includegraphics[width=1\columnwidth]{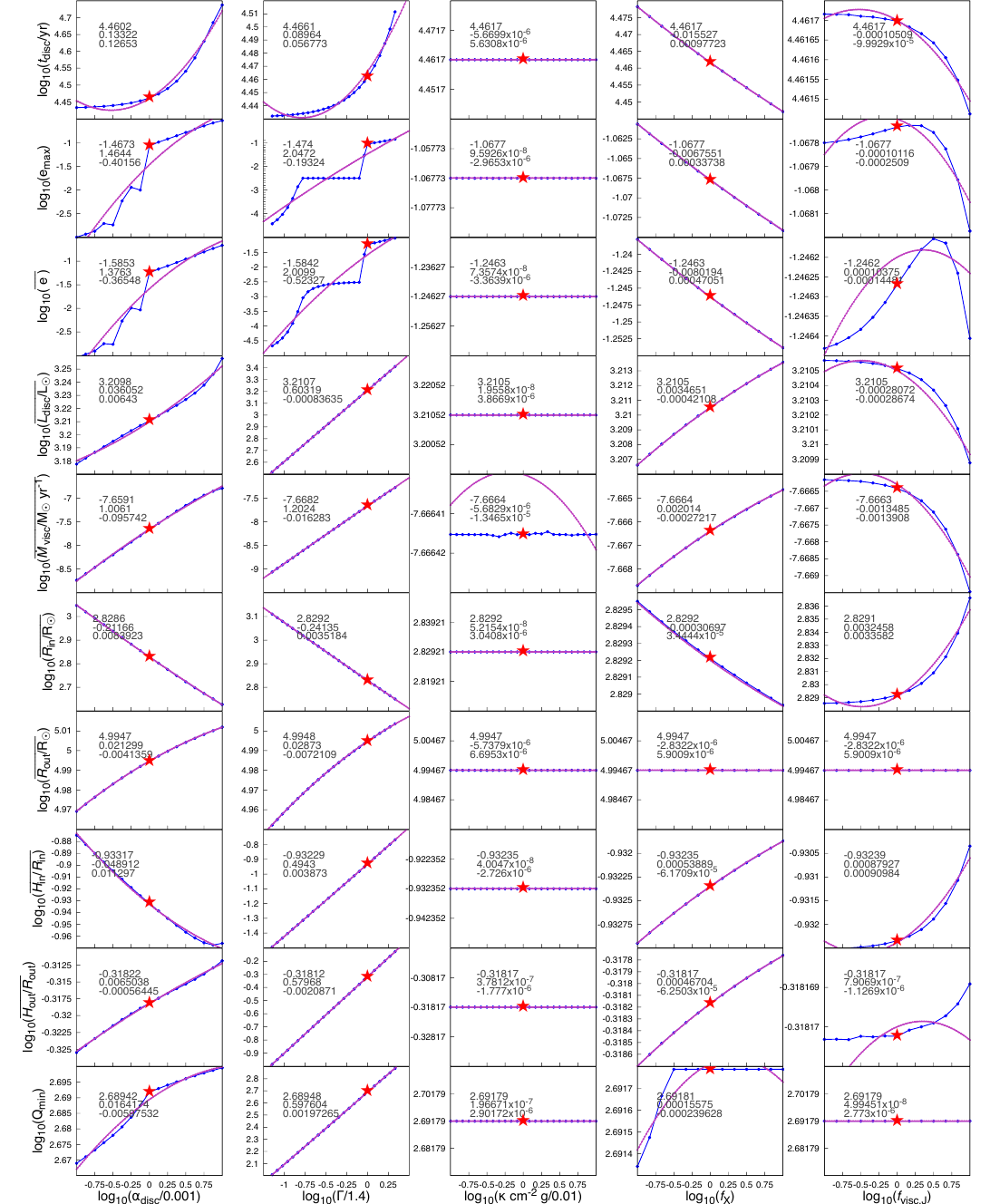}
\par\end{centering}
\caption{\label{fig:fits1}Blue lines show the logarithm of our \IRAS{} disc model properties,
$\log_{10}\left(y\right)$ where $y$ is (top to bottom) disc lifetime $\tdisc$, maximum eccentricity $\emax$, mean eccentricity $\overline{e}$, disc luminosity $\overline{\Ldisc}$, inner-edge mass inflow rate $\overline{\Mdotvisc}$, disc inner and outer edge radii $\overline{\rin}$ and $\overline{\rout}$, scale height ratios at the inner and outer edges $\overline{\hin/\rin}$ and $\overline{\hout/\rout}$ and minimum Toomre $Q$ parameter as functions of logarithmic disc parameters, $\log\left(x/x_{0}\right)$: viscosity parameter $\adisc$, gas adiabatic index $\Gamma$, disc opacity $\kappa$, X-ray disc wind multiplier $\fX$ and inflow viscosity multiplier $\fviscJ$, where $x_{0}$ is the default parameter shown by the red stars. Overbars indicate values time-averaged over the life of the disc.
The coefficients labelled in each plot, from top to bottom, are $\Phi$, $\Upsilon$
and $\Theta$ in the fitting function $\log_{10}\left(y\right)=\Phi+\Upsilon\log_{10}\left(x/x_{0}\right)+\Theta\left[\log_{10}\left(x/x_{0}\right)\right]^{2}$
shown in magenta.}
\end{figure}

\begin{figure}
\centerfloat
\begin{centering}
\includegraphics[width=1\columnwidth]{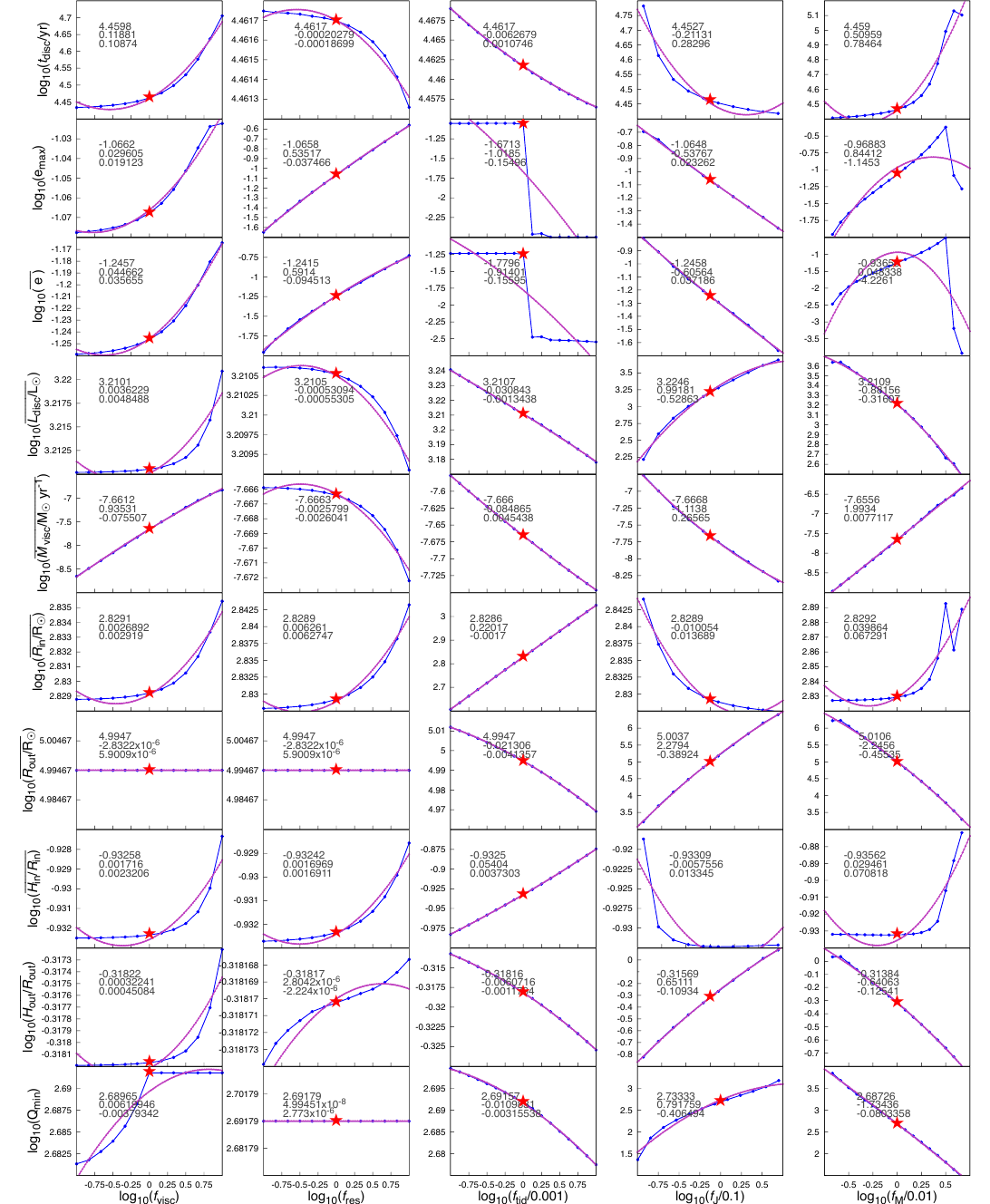}
\par\end{centering}
\caption{\label{fig:fits2}As \figref~\ref{fig:fits1} but as functions of the viscous-inflow parameter $\fvisc$, eccentricity-pumping resonance factor $\fres$, tidal-torque parameter $\ftid$, disc initial angular momentum parameter $\fJ$ and disc initial mass parameter $\fM$.}
\end{figure}

In Figs.~\ref{fig:fits1} and~\ref{fig:fits2}
we fit observational properties, $y$, of our circumbinary-disc model to the model parameters, $x$ (Table~\ref{tab:disc-model-parameters}). In what follows, overbars indicate properties time-averaged
over the lifetime of the disc. The disc properties, $y$, shown from
top to bottom are the logarithm of the minimum Toomre $Q$ over the
life of the disc, the logarithm of the average ratio of the disc scale
height at its outer radius to its outer radius, $\log_{10}\left(\overline{\hout/\rout}\right)$,
the logarithm of the average ratio of the disc scale height at its
inner radius to its inner radius, $\log_{10}\left(\overline{\hin/\rin}\right)$,
the outer and inner average radii, $\log_{10}\left(\overline{\rout/\mathrm{R_{\odot}}}\right)$
and $\log_{10}\left(\overline{\rin/\mathrm{R_{\odot}}}\right)$ respectively,
the average disc mass, $\log_{10}\left(\overline{\mdisc/\mathrm{M_{\odot}}}\right)$,
the average disc luminosity, $\log_{10}\left(\overline{L_{\mathrm{disc}}/\mathrm{L_{\odot}}}\right)$,
the average binary eccentricity, $\log_{10}\left(\overline{e}\right)$,
the maximum binary eccentricity, $\log_{10}\left(e_{\mathrm{max}}\right)$
and the disc lifetime, $\log_{10}\left(\tdisc/\mathrm{yr}\right)$.
The disc parameters, $x$, are shown as $\log_{10}\left(x/x_{0}\right)$
where $x_{0}$ is our default value of the parameter. We then fit
the observable, $\log_{10}\left(y\right)$, to a quadratic function
in $\log_{10}\left(x/x_{0}\right)$ such that,
\begin{eqnarray}
\log_{10}\left(y\right) & = & \Phi+\Upsilon\log_{10}\left(x/x_{0}\right)+\Theta\left[\log_{10}\left(x/x_{0}\right)\right]^{2}\,,\label{eq:polyfit}
\end{eqnarray}
where $\Phi$, $\Upsilon$ and $\Theta$ are the coefficients shown
in each plot. 

In many cases, there is little variation of $y$ with $x$, hence
$\Upsilon$ and $\Theta$ are small. Our fits are poor when the data
bifurcates, e.g.~$\log_{10}\left(e_{\mathrm{max}}\right)$ vs $\log_{10}\left(\adisc\right)$
or $\log_{10}\left(\ftid\right)$, and in these cases the reader should
take care to check Figs.~\ref{fig:fits1} and~\ref{fig:fits2}. 

\vfill{}

\pagebreak{}

\section{MCMC results}

\label{sec:MCMC-results}
\begin{figure}
\begin{centering}
\includegraphics{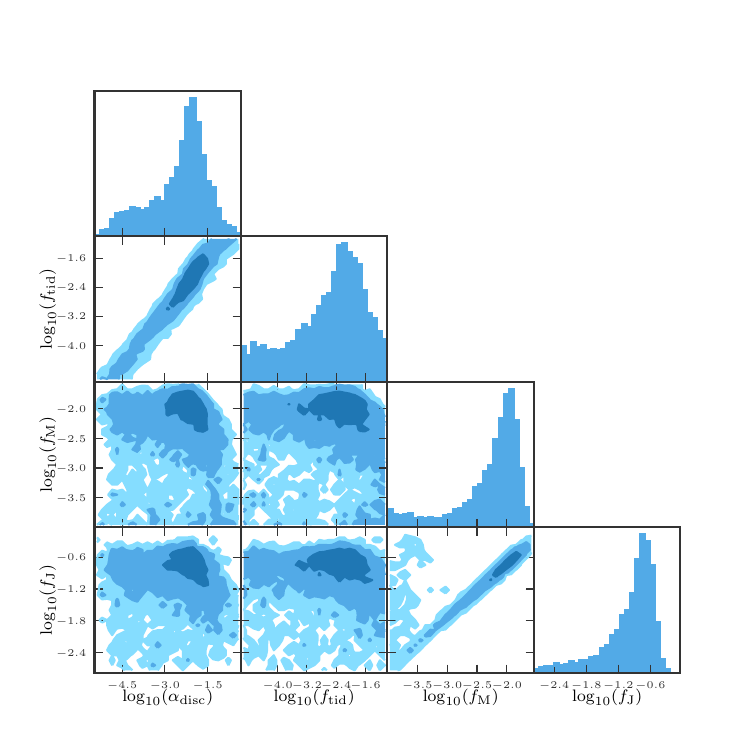}
\par\end{centering}
\caption{\label{fig:gtc}Log-likelihood as a function of model parameters $\protect\fM$,
$\protect\fJ$, $\protect\ftid$ and $\protect\adisc$ fitted to observations
of $\protect\IRAS$ combined with the extra constraints $\protect\Mdotvisc=10^{-7}\pm0.5\times10^{-7}\mathrm{\,M_{\odot}}\,\protect\yr^{-1}$
and $\protect\rin/a=2.00\pm0.25$. Shades of blue, from darkest to
lightest, represent the two-dimensional $1\sigma$, $2\sigma$ and
$3\sigma$ contours.}
\end{figure}

\begin{figure}
\begin{centering}
\includegraphics{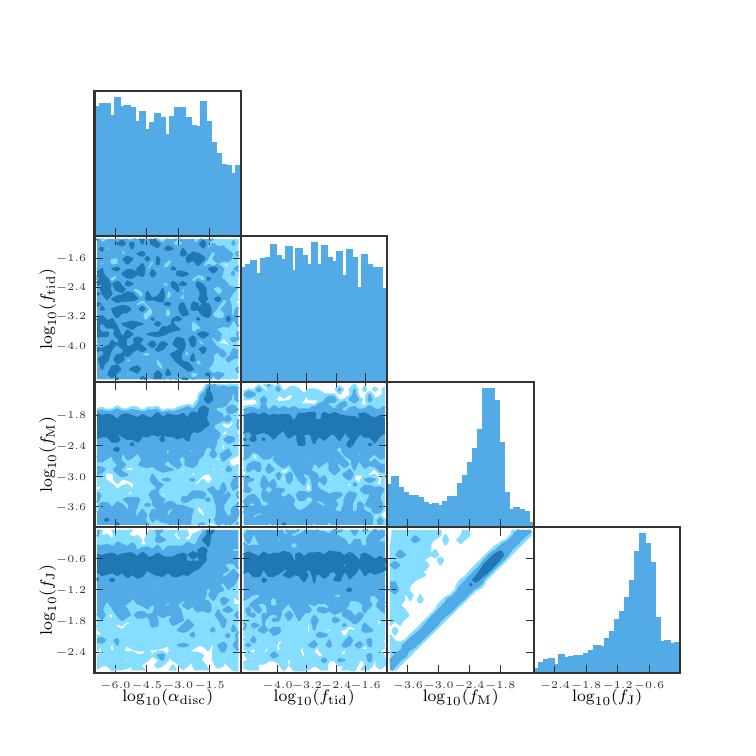}
\par\end{centering}
\caption{\label{fig:gtc-no_Mvisc-no_Rina}Log-likelihood as a function of model
parameters $\protect\fM$, $\protect\fJ$, $\protect\ftid$ and $\protect\adisc$
fitted to observations of $\protect\IRAS$ without conditions on $\protect\Mdotvisc$
and $\protect\rin/a$. Shades of blue, from darkest to lightest, represent
the two-dimensional $1\sigma$, $2\sigma$ and $3\sigma$ contours.}
\end{figure}

Figs.~\ref{fig:gtc} and \ref{fig:gtc-no_Mvisc-no_Rina}
show the log-likelihood as a function of model parameters $\fM$,
$\fJ$, $\ftid$ and $\adisc$ calculated by comparing our time-averaged
disc model to the observations of $\IRAS$ combined with and without
conditions on $\Mdotvisc$ and $\rin/a$, respectively, as described
in \secref{}~\ref{subsec:Best-fitting-disc-parameters}. Our Giant-Triangle-Confusograms
were created using the $\code{pygtc}$ package \citep{Bocquet2016}.


\section{Solution accuracy checks}
\label{sec:solution-accuracy-checks}
\begin{figure}
\begin{centering}
\includegraphics{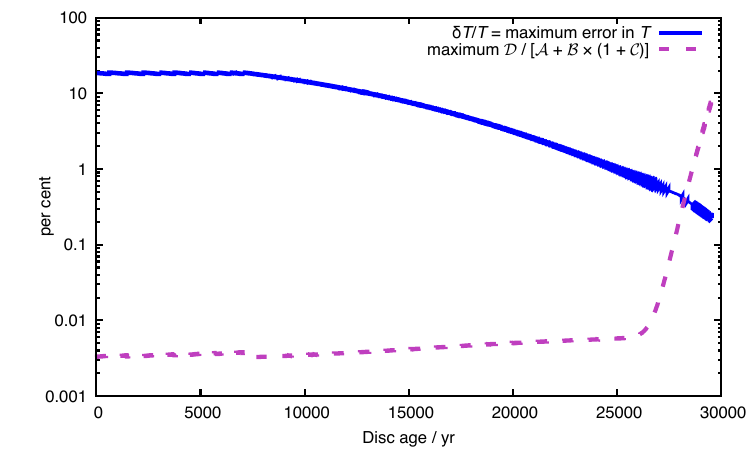}
\par\end{centering}
\caption{\label{fig:solution-accuracy-checks}
  Maximum error in our \IRAS{} circumbinary-disc temperature solution, $\delta T/\,\f{T}{R}$ (blue, solid line; \eqq{}~\ref{eq:deltaT}), and maximum contribution from the mass-loss term, $\mathcal{D}$, relative to $\mathcal{A} + \mathcal{B}(1+\mathcal{C})$ (magenta, dashed line). At each disc age, the maximum is computed from the values throughout the disc.
}
\end{figure}

Our solution method (Secs.~\ref{subsec:Disc-structure} and \ref{subsec:Solution-method}) has a maximum theoretical error in disc temperature, $T(R)$, of $32\,\pc$.
We estimate this error in our \IRAS{} disc solution by rearranging \eqq{}~\ref{eq:disc-heat-balance-1} to,
\begin{equation}
\delta T = \sigma T^{4} - \mathcal{A} - \mathcal{B}(1+\mathcal{C})\,,\label{eq:deltaT}
\end{equation}
and show $\delta T/T$ over the lifetime of our \IRAS{} model in \figref~\ref{fig:solution-accuracy-checks}. The magnitude of $\delta T/T$ is always less than $20\,\pc$.

In \figref~\ref{fig:solution-accuracy-checks} we also show
$\mathcal{D} / \big[\mathcal{A} + \mathcal{B}(1+\mathcal{C})\big]$ to estimate the magnitude of the neglected mass-loss term, $\mathcal{D}$, which for most of the evolution is negligible, and even near the end of the disc's life when mass loss is greatest, the $\mathcal{D}$ term never exceeds $10\,\pc$.

\end{document}